\documentclass[a4paper,10pt]{article}
\usepackage[utf8x]{inputenc}
\usepackage{amsmath}
\usepackage{amssymb}
\usepackage{amsthm}
\usepackage{slashed}
\usepackage{upgreek}
\usepackage{mathrsfs}
\usepackage{bbm}
\numberwithin{equation}{section}

\let\originalleft\left
\let\originalright\right
\renewcommand{\left}{\mathopen{}\mathclose\bgroup\originalleft}
\renewcommand{\right}{\aftergroup\egroup\originalright}

\newcommand{\bigindent}{\qquad\qquad}
\newcommand{\reverseindent}{\mkern-72mu}

\def\bbe{{\bf{e}}}
\font\mybb=msbm10 at 11pt

\def\bb#1{\hbox{\mybb#1}}

\def\bR {\bb{R}}

\def\bN {\bb{N}}

\newcommand{\bea}{\begin{eqnarray}}
\newcommand{\eea}{\end{eqnarray}}

\begin{document}

\begin{titlepage}
\begin{center}
\vspace*{-1.0cm}
\hfill DMUS--MP--15/03 \\

\vspace{2.0cm} {\Large \bf Supersymmetry of IIA warped flux AdS and flat backgrounds } \\[.2cm]

\vskip 2cm
S. Beck$^1$,  J. B.  Gutowski$^2$ and G. Papadopoulos$^1$
\\
\vskip .6cm

\begin{small}
$^1$\textit{  Department of Mathematics, King's College London
\\
Strand, London WC2R 2LS, UK.
\\
E-mails: samuel.beck@kcl.ac.uk,
\\
george.papadopoulos@kcl.ac.uk}
\end{small}\\*[.6cm]

\begin{small}
$^2$\textit{Department of Mathematics,
University of Surrey \\
Guildford, GU2 7XH, UK \\
Email: j.gutowski@surrey.ac.uk}
\end{small}\\*[.6cm]

\end{center}

\vskip 3.5 cm
\begin{abstract}

We identify the fractions of supersymmetry preserved by the most general warped flux AdS and flat backgrounds in both massive and
standard IIA supergravities. We find that $AdS_n\times_w M^{10-n}$ preserve $2^{[{n\over2}]} k$ for $n\leq 4$ and $2^{[{n\over2}]+1} k$ for $4<n\leq 7$ supersymmetries, $k\in \bN_{>0}$. In addition we show that, for suitably restricted fields and
$M^{10-n}$, the killing spinors of AdS backgrounds are given in terms
of the zero modes of Dirac like operators on $M^{10-n}$. This generalizes the Lichnerowicz  theorem
for connections whose holonomy is included in a general linear group. We also adapt our results
to $\bR^{1,n-1}\times_w M^{10-n}$ backgrounds which underpin flux compactifications to $\bR^{1,n-1}$  and show that these preserve
$2^{[{n\over2}]} k$ for $2<n\leq 4$, $2^{[{n+1\over2}]} k$ for $4<n\leq 8$, and $2^{[{n\over2}]} k$ for $n=9, 10$ supersymmetries.

\end{abstract}

\end{titlepage}

\section{Introduction}

AdS backgrounds have found widespread applications in supergravity compactifications and more recently
in AdS/CFT correspondence, see reviews \cite{duff, grana, maldacena}. Because of this there is an extensive literature on the subject starting from the original work of \cite{FR}, for some selected publications on AdS backgrounds in the context of IIA supergravity see \cite{romans}-\cite{passias}. A first step towards the classification of AdS backgrounds is to identify the fractions of
supersymmetry that are  preserved. This has been established for D=11 and IIB  AdS backgrounds in
\cite{mads} and \cite{iibads}, respectively. The novelty of this approach is that no restrictions have been put on either the form of the fields or that
of the Killing spinors. As a result the most general warped flux AdS backgrounds have been considered. It has also been found that the Killing spinors
do not factorize into Killing spinors on the AdS and Killing spinors on the transverse space.

One of the aims of this paper is to  count the number of supersymmetries of warped flux AdS backgrounds
in both  standard \cite{huq, petrini, west} and massive \cite{romans}  IIA supergravities, and so complete this analysis for all D=11 and type II
supergravities. In what follows, we restrict the spacetime
to be a warped product of a $AdS_n$ space with a transverse space $M^{10-n}$  and require that the fluxes
 respect the isometries of $AdS_n$ . We do not place any other assumptions on either the fluxes or on the form of Killing spinors. We find that such  warped flux $AdS_n\times_w M^{10-n}$
backgrounds\footnote{These backgrounds are taken up to discrete identifications.} preserve
\bea
N= 2^{[{n\over2}]} k~,~~~n\leq 4~;~~~ N=2^{[{n\over2}]+1} k~,~~~4<n\leq 7~,
\label{cadssusy}
\eea
supersymmetries, where  $k\in \bN_{>0}$. This formula gives the a priori number of supersymmetries preserved. It is expected
that there are additional restrictions on $N$.  For example $N<32$ as there are no (massive) IIA  $AdS_n\times_w M^{10-n}$ backgrounds which are maximally supersymmetric \cite{maxsusy}.
The proof  that $AdS_2\times_w M^8$ backgrounds preserve an even number of supersymmetries
requires the additional assumption that $M^8$ and the fields satisfy suitable conditions
such that the maximum principle applies, eg $M^8$ is closed and fields are smooth. The result is a special case of the more general theorem that
all near horizon geometries of (massive) IIA supergravity preserve an even number of supersymmetries
given in \cite{iiahor}. For the counting of supersymmetries for the rest of $AdS_n\times_w M^{10-n}$, $n>2$, backgrounds no such assumption is necessary.
A summary of these results is also presented in table 1.

Furthermore, we show that the Killing spinors of the $AdS_n\times_w M^{10-n}$ backgrounds can be determined from the zero modes of Dirac-like operators on $M^{10-n}$ which depend on the fluxes.
For this we demonstrate new Lichnerowicz type theorems, using the maximum principle, which relate the
Killing spinors to the zero modes of these Dirac-like operators.

In the limit that the AdS radius goes to infinity, the $AdS_n\times_w M^{10-n}$ backgrounds
become the most general warped flux flat backgrounds $\bR^{n-1,1}\times_w M^{10-n}$. The latter have also widespread applications in supergravity, string theory and M-theory
as they include the most general flux compactifications. Taking the limit of infinite AdS radius we adapt most of our $AdS_n\times_w M^{10-n}$ results to $\bR^{n-1,1}\times_w M^{10-n}$
backgrounds. In particular, all our local computations are valid in this limit and so one can establish that the number of supersymmetries preserved by such backgrounds are
\bea
&&N=2^{[{n\over2}]} k~,~~2<n\leq 4~; ~~N=2^{[{n+1\over2}]} k~,~~~4<n\leq 8~;~
\cr
&&N=N=2^{[{n\over2}]} k~,~~n=9, 10~,
\label{cafsusy}
\eea
where $k\in\bN_{>0}$. There are additional restrictions on $N$.  In particular it is known that the  maximally supersymmetric solutions of standard
IIA supergravity are locally isometric to Minkowski spacetime, the fluxes vanish and the dilaton is constant, and moreover that the massive IIA supergravity does not have
a maximally supersymmetric solution  \cite{maxsusy}. In addition, if the Killing spinors do not depend on the $\bR^{n-1,1}$ coordinates, then all backgrounds with $N>16$ are locally
 isometric to $\bR^{9,1}$ with zero  fluxes and constant dilaton as a consequence of the homogeneity conjecture \cite{josehom}. These results have also been collected in table 2.
Note that the counting of supersymmetries in AdS and flat backgrounds is different.  This is because there are differences in the counting
of linearly independent  Killing spinors for finite and infinite AdS radius.

Apart from the similarities  there are also some differences between $AdS_n\times_w M^{10-n}$  and $\bR^{n-1,1}\times_w M^{10-n}$ backgrounds. First, some of the regularity results
that have  used to prove the new Lichnerowicz   type theorems for $AdS_n\times_w M^{10-n}$   are no longer valid
for $\bR^{n-1,1}\times_w M^{10-n}$ backgrounds. This is related to the property that flux compactifications to $\bR^{n-1,1}$  without higher order corrections,
or without the addition of sources, are all singular \cite{maldacena2}.
 As a consequence, it is not possible to prove that  $\bR^{1,1}\times_w M^{8}$ backgrounds preserve an even number of supersymmetries, and so there is no a priori restriction on the number of supersymmetries preserved by such backgrounds; though $N\not=31$ because of the classification result of \cite{bandos}. In addition, it is not
 straightforward to adapt the proof of new Lichnerowicz   type theorems from  $AdS_n\times_w M^{10-n}$
 to $\bR^{n-1,1}\times_w M^{10-n}$ backgrounds. Even though   the formulae used for the application of the maximum principle are still valid for flux $\bR^{n-1,1}\times_w M^{10-n}$ backgrounds, the fields violate the regularity
 assumptions which are necessary for the application of the maximum principle.

Our analysis also reveals that the Killing spinors of $AdS_n\times_w M^{10-n}$ spaces do not factorize into the Killing spinors on  $AdS_n$ and Killing spinors
on $M^{10-n}$. This result is similar to that already established for the D=11 and IIB backgrounds in \cite{mads, iibads} where it was shown that such a factorization
leads to an incorrect counting of Killing spinors. Similar results also hold for $\bR^{n-1,1}\times_w M^{10-n}$ backgrounds.

To prove these results, we first solve the KSEs of (massive) IIA supergravity along the AdS directions  without assuming a special form for the Killing spinor.
A convenient way to do this is to write these backgrounds as near horizon geometries as suggested in \cite{adshor} and then we use the results of \cite{iiahor}. For warped flux $AdS_2\times_w M^8$ backgrounds, the counting of supersymmetries and the rest of the results are  a special case of those of \cite{iiahor} for IIA horizons. For the rest of the $\mathrm{AdS}_n\times_w M^{10-n}$
backgrounds, we  integrate the KSEs along the $AdS_n$ directions and demonstrate that the Killing spinors $\epsilon$ depend on the coordinates of $\mathrm{AdS}_n$ and four 16-component spinors $\sigma_\pm, \tau_\pm$ which in turn
depend only on the coordinates of $M^{10-n}$. Moreover, the process of integration over $\mathrm{AdS}_n$ introduces one new  algebraic KSE
for each $\sigma_\pm, \tau_\pm$.
 Thus each spinor $\sigma_\pm, \tau_\pm$ obeys
now three KSEs, one parallel transport equation which arises from the gravitino KSE
of (massive) IIA supergravity, one algebraic KSE which arises from the dilatino KSE of (massive) IIA supergravity, and the new algebraic KSE.
The counting of supersymmetries then proceeds with the observation that there are Clifford algebra
operators which intertwine between the triplets of KSEs.  As a result,
given a solution in one triplet of KSEs, these Clifford algebra operators generate solutions in the other triplets of KSEs. Counting the linearly independent solutions generated this way one proves (\ref{cadssusy}). The proof of (\ref{cafsusy}) for $\bR^{n-1,1}\times_w M^{10-n}$ backgrounds is similar.

The proof of the correspondence of Killing spinors and zero modes of Dirac-like operators ${\mathscr D}^{(\pm)}$ on $M^{10-n}$ for $\mathrm{AdS}_n\times_w M^{10-n}$ relies on the application
of the maximum principle. First a suitable choice is made for  ${\mathscr D}^{(\pm)}$ as there are several options available because of the presence of algebraic
KSEs. Then assuming that a spinor $\chi_+$ is a zero mode of ${\mathscr D}^{(+)}$ and after using the field equations and Bianchi identities, one schematically establishes
\bea
\nabla^2\parallel\chi_+\parallel^2- \alpha^i \nabla_i \parallel\chi_+\parallel^2={\cal Q} (\nabla^{(+)}\chi, {\cal A}^{(+)}\chi,  \mathbb{B}^{(+)}\chi_+) \geq 0~,
\label{maxprii}
\eea
where $\chi_+= \sigma_+$ or $\tau_+$, and ${\cal Q}$ is a function
which vanishes if and only if the triplet of KSEs  $\nabla^{(+)}\chi_+=0$, ${\cal A}^{(+)}\chi_+=0$ and $\mathbb{B}^{(+)}\chi_+=0$
is satisfied. ${\cal Q}$ is specified in each case. An application of the maximum principle reveals that the only solution to the above equation is that $\parallel\chi_+\parallel^2$ is constant and that $\chi_+$
satisfies the KSEs. A similar formula can be established for the $\sigma_-$ and $\tau_-$ spinors.

This paper has been organized as follows.  In sections 2, 3, 4, and 5, we present the proof of the formula (\ref{cadssusy}) for all $\mathrm{AdS}_n\times_w M^{10-n}$ backgrounds,
and demonstrate the new Lichnerowicz   type theorems. In section 6, we present the proof for the formula (\ref{cafsusy}) for all $\bR^{n-1,1}\times_w M^{10-n}$ backgrounds.
In section 7, we examine the factorization properties of the Killing spinors for $\mathrm{AdS}_n\times_w M^{10-n}$  and $\bR^{n-1,1}\times_w M^{10-n}$ backgrounds, and in section
8 we give our conclusions. In appendix A, we state our conventions and in appendices B, C, and D, we prove the formula (\ref{maxprii}) for $\mathrm{AdS}_n\times_w M^{10-n}$, $2<n\leq 7$,  backgrounds.

\section{$\mathrm{AdS}_2\times_w M^8$}

\subsection{Fields, Bianchi identities and Field Equations}

\subsubsection{Fields}

As  has already been mentioned, all AdS backgrounds are included in the near horizon geometries. To describe the fields
of $\mathrm{AdS}_2\times_w M^8$ it suffices to impose the isometries of the $AdS_2$ space on all the fields of the near horizon
geometries of \cite{iiahor}.  In such a case, the fields\footnote{The choice of the fields of $\mathrm{AdS}_2\times_w M^8$ backgrounds here
is different from that of near horizon geometries in \cite{iiahor}. In particular all R-R fields have been multiplied by $e^\Phi$. For more details see   \cite{howe} and \cite{iiaclass}
and appendix A.}  can be written as
\bea
ds^2&=&2 \bbe^+ \bbe^- + ds^2(M^8)~,
\cr
G&=& \bbe^+ \wedge \bbe^-\wedge X+ Y~,~~~H=\bbe^+ \wedge \bbe^-\wedge W+ Z~,~~~
\cr
F&=&\bbe^+ \wedge \bbe^- N+ P~,~~~S=S~,~~~\Phi=\Phi~.
\eea
where $X$ and $P$ are 2-forms on $M^8$, $Y$ is a 4-form on $M^8$, $Z$ is a 3-form on $M^8$, and $N$ and the dilaton $\Phi$ are functions on $M^8$.
$S=e^\Phi m$, where $m$ is the mass parameter of massive IIA supergravity. For the standard IIA supergravity $m=0$ and so $S=0$. Furthermore,
\bea
\bbe^+&=&du~,~~~\bbe^-=(dr+ rh-{1\over2} r^2 \Delta du)~,~~~
\cr
h&=&-2 A^{-1} dA=\Delta^{-1} d\Delta~,~~~\Delta= \ell^{-2} A^{-2}~,
\eea
where the dependence on the coordinates $u,r$ is explicit, $A$ is the warp factor which depends only on the coordinates of $M^8$ and $\ell$ is the radius of $\mathrm{AdS}_2$.

\subsubsection{Bianchi identities and Field equations}

The Bianchi identities of (massive) IIA supergravity reduce to differential identities on the components of the fields localized on $M^8$.
In particular a direct computation reveals that
\bea
d(A^2 W)&=&0~,~~~d(A^2 X)-A^2  d\Phi\wedge X- A^2 W\wedge P- A^2 N Z=0~,~~~
\cr
dZ&=&0~,~~~d (A^2 N)-A^2 N d\Phi - S A^2 W=0~,~~~
\cr
dY-d\Phi\wedge Y&=&Z\wedge P~,~~~dP-d\Phi \wedge P=S Z~.
\eea
Similarly, the field equations of the (massive) IIA supergravity decomposed as
\bea
&&\nabla^j P_{ji}+(2\partial^j \log A-\partial^j \Phi) P_{ji}-W^j X_{ji}+{1\over6} Z^{jk\ell} Y_{jk\ell i}=0~,
\cr
&&e^{2\Phi} \nabla^i(e^{-2\Phi} W_i)- S N-{1\over2} P^{ij} X_{ij}+{1\over48}  *Y_{i_1\dots i_4} Y^{i_1\dots i_4}=0~,
\cr
&&e^{2\Phi}\nabla^k(e^{-2\Phi} Z_{kij})- S P_{ij}+ 2 \partial^k\log A Z_{kij}+ N X_{ij}-{1\over2} P^{kl} Y_{klij}
\cr&&~~~~-{1\over2}  X_{k\ell}\, *Y_{ij}{}^{k\ell}=0~,
\cr
&&\nabla^j X_{ji}-\partial^j\Phi X_{ji}+{1\over6}  *Y_i{}^{ k_1k_2 k_3}  Z_{k_1k_2k_3}=0~,
\cr
&&\nabla^i Y_{ijk\ell}+ (2\partial^i\log A-\partial^i\Phi) Y_{ijk\ell}-{1\over2}  X_{m_1m_2} \,*Z_{jk\ell}{}^{m_1m_2}
\cr&&~~~~
-   *Y_{jk\ell}{}^{n} W_n=0~,
\cr
&&\nabla^2\Phi+ 2A^{-1} \partial^iA \partial_i\Phi= 2 \partial^i\Phi \partial_i\Phi+{1\over 2} W^2-{1\over12} Z^2-{3\over4} N^2
\cr
&&~~~+{3\over8} P^2-{1\over8} X^2+{1\over 96} Y^2+{5\over4} S^2~,
\eea
and in particular the Einstein equation decomposes as
\bea
&&\nabla^i\partial_i \log A+\Delta+2 (d\log A)^2=2  \partial^i \log A \,\partial_i \Phi+{1\over2} W^2
\cr&&~~~~+{1\over4} N^2+{1\over8}X^2+
{1\over8} P^2-{1\over96} Y^2-{1\over4} S^2~,
\cr
&&R^{(8)}_{ij}=2\nabla_{i} \partial_{j}\log  A+2  \partial_i\log A \partial_j\log A-2 \nabla_i\partial_j\Phi-{1\over2} W_i W_j+{1\over4} Z^2_{ij}
+{1\over2}P^2_{ij}
\cr
&&-{1\over2} X^2_{ij} +{1\over12} Y^2_{ij} +
\delta_{ij} \big({1\over4} N^2-{1\over4} S^2-{1\over8} P^2+{1\over8} X^2-{1\over96} Y^2\big)~,
\eea
where $\nabla$ is the Levi-Civita connection on $M^8$ and the  Latin indices $i,j,k, \dots$ are frame $M^8$ indices.

\subsection{Local aspects: Solutions of KSEs}

\subsubsection{Solution of KSEs along $\mathrm{AdS}_2$}

The solution of the KSEs for $\mathrm{AdS}_2\times_w M^8$ backgrounds is a special case of that presented for IIA horizons in \cite{iiahor}.   In particular, the solution of the KSEs along the $\mathrm{AdS}_2$ directions
can be written as
\bea
&&\epsilon=\epsilon_++ \epsilon_-~,~~~
\cr
&&\epsilon_+=\eta_++u\Gamma_+\Theta_-\eta_-~,~~~\epsilon_-=\eta_-+ r\Gamma_- \Theta_+ \big(\eta_++u\Gamma_+\Theta_-\eta_-\big)~,
\label{ksesolads2}
\eea
where $\Gamma_\pm\epsilon_\pm=0$,
\bea
\Theta_\pm=-{1\over2} A^{-1} \slashed{\partial} A\mp\Gamma_{11}  \slashed{W}-{1\over16} \Gamma_{11} (\pm 2 N+  \slashed{P})-{1\over 8 \cdot 4!} (\pm 12  \slashed{X}+ \slashed{Y})-{1\over8} S~,
\eea
and $\eta_\pm$ depend only on the coordinates of $M^8$. This summarizes the solution of the KSEs along the $\mathrm{AdS}_2$ directions.

\subsubsection{Independent KSEs on $M^8$}

Having solved the KSEs along the $\mathrm{AdS}_2$ directions, it remains to identify the remaining independent KSEs.
This is not straightforward. After substituting (\ref{ksesolads2}) back into the KSEs of (massive) IIA supergravity and  expanding in the $u$ and $r$ coordinates,
one finds  a large number of conditions. These can be interpreted as integrability conditions along the  $\mathrm{AdS}_2$ and mixed $\mathrm{AdS}_2$  and $M^8$ directions.
However after an extensive analysis which involves  the use  of  Bianchi identities and field equations,
one finds that the remaining independent KSEs are
\bea
\nabla^{(\pm)}_i \eta_\pm=0~,~~~{\cal A}^{(\pm)} \eta_\pm=0~,
\label{kseadsm8}
\eea
where
\bea
\nabla^{(\pm)}_i=\nabla_i+ \Psi^{(\pm)}_i~,
\eea
and
\bea
\Psi^{(\pm)}_i&=&\pm{1\over2} A^{-1} \partial_iA\mp{1\over 16} \slashed{X} \Gamma_i+{1\over 8 \cdot 4!} \slashed{Y} \Gamma_i+{1\over8} S \Gamma_i
\cr
&&~~~
+\Gamma_{11} \big(\mp{1\over4} W_i+{1\over8} \slashed{Z}_i \pm{1\over8} N \Gamma_i-{1\over16} \slashed{P} \Gamma_i\big)~,
\eea
and
\bea
{\cal A}^{(\pm)}&=&\slashed{\partial} \Phi+\big(\mp {1\over8} \slashed{X}+{1\over 4\cdot 4!} \slashed{Y}+{5\over4} S\big)
\cr
&&~~~+\Gamma_{11} \big(\pm {1\over2} \slashed{W}-
{1\over12}\slashed{Z}\mp {3\over4} N+{3\over8} \slashed{P}\big)~.
\eea
Furthermore, one can show that if $\eta_-$ is a Killing spinor, ie satisfies (\ref{kseadsm8}), then
\bea
\eta_+=\Gamma_+ \Theta_-\eta_-~,
\eea
is also a Killing spinor.

\subsubsection{Counting supersymmetries}

The investigation so far is not sufficient to prove that the number of supersymmetries preserved by $\mathrm{AdS}_2\times_w M^8$ backgrounds is even, as given in (\ref{cadssusy}).
To prove this, some additional restrictions on the backgrounds are necessary which will be described in the next section.

\subsection{Global aspects: Lichnerowicz type theorems}

\subsubsection{The non-vanishing of warp factor $A$}

To proceed, we shall show that if $A$ and the fields are smooth, then $A$ does not vanish on $M^8$.  The argument which proves this is similar to that used
in \cite{mads} and \cite{iibads} to demonstrate the analogous statements for D=11 and IIB AdS backgrounds, and where a more detailed analysis is presented. Here
we present a brief description of the proof which relies on the field equation of  $A$. Assuming that $A$ does not vanish everywhere on $M$,
we multiply that field equation of $A$  with $A^2$ at a value for which $A^2\not=0$ to find
\bea
&&-A \nabla^i \partial_i A-\ell^{-2}- \partial^i A \partial_i A=-2 A \partial^i A \partial_i \Phi-{1\over2}A^2 W^2-{1\over4} A^2N^2
\cr
&&~~~~-{1\over8}A^2 X^2-
{1\over8} A^2 P^2-{1\over96}A^2 Y^2-{1\over4} A^2S^2~.
\eea
Then taking a sequence that converges to a point in $M^8$ that $A$ vanishes, we find that if such a point exists it is inconsistent with the above
field equation as $\ell$ is the radius of $\mathrm{AdS}_2$ which is finite.  As a result for smooth solutions, $A$ cannot vanish anywhere
on $M^8$.

\subsubsection{Lichnerowicz type theorems for $\eta_\pm$}

The Killing spinors $\eta_\pm$ can be identified with the zero modes of a suitable Dirac-like operator on $M^8$. In particular, let us
define
\bea
{\mathscr D}^{(\pm)}=\slashed{\nabla}^{(\pm)} -{\cal A}^{(\pm)}~,
\eea
where
$\slashed{\nabla}^{(\pm)}=\slashed{\nabla}+\Psi^{(\pm)}$, $\slashed{\nabla}$ is the Dirac operator on $M^8$, and
\bea
\Psi^{(\pm)}\equiv \Gamma^i \Psi^{(\pm)}_i=\pm{1\over2} A^{-1}\slashed{\partial} A\mp{1\over4}\slashed{X}+ S+\Gamma_{11} \big(\pm{1\over4} \slashed{W}-{1\over8} \slashed Z\mp N+{1\over4} \slashed{P}\big). \
\eea
It turns out that if the fields and $M^8$ satisfy the requirements for the maximum principle to apply, eg $M^8$ is compact without boundary and
all the fields are smooth, then
\bea
\nabla_i^{(\pm)}\eta_\pm=0~,~~~{\cal A}^{(\pm)}\eta_\pm=0\Longleftrightarrow {\mathscr D}^{(\pm)}\eta_\pm=0~.
\eea
It is clear that the proof of this in the forward direction is straightforward. To establish the opposite direction for the $\eta_+$ spinors, let us assume that
${\mathscr D}^{(+)}\eta_+=0$. Then after some extensive algebra  using the Bianchi identities and the field equations, one finds \cite{iiahor} that
\bea
&&\nabla^2\parallel\eta_+\parallel^2-2(\partial^i\Phi-A^{-1}\partial^iA)\nabla_i\parallel\eta_+\parallel^2=
\cr&&~~~~~~~2 \parallel\hat\nabla^{(+)}\eta_+\parallel^2
- (4\kappa+16\kappa^2) \parallel{\cal A}^{(+)}\eta_+\parallel^2~,
\eea
where
\bea
\hat\nabla_i^{(\pm)}=\nabla_i^{(\pm)}+\kappa \Gamma_i {\cal A}^{(\pm)}~.
\eea
Applying the maximum principle for $\kappa\in(-{1\over4}, 0)$,  one concludes that the solutions of the above equation are  Killing spinors and that
\bea
\parallel\eta_+\parallel=\mathrm{const}~.
\eea

Similarly assuming that ${\mathscr D}^{(-)}\eta_-=0$, one can establish the identity
\bea
&&\nabla^2\big (A^{-2} \parallel \eta_-\parallel^2\big)-2(\partial^i\Phi-A^{-1}\partial^iA)\nabla_i (A^{-2} \parallel\eta_-\parallel^2)=
\cr&& ~~~~~~
2 A^{-2} \parallel\hat\nabla^{(-)}\eta_-\parallel^2- (4\kappa+16\kappa^2) A^{-2} \parallel{\cal A}^{(-)}\eta_-\parallel^2~.
\eea
Again the application of the maximum principle for $\kappa\in(-{1\over4}, 0)$ gives that $\eta_-$ is a Killing spinor and that
\bea
A^{-1} \parallel \eta_-\parallel=\mathrm{const}~.
\eea
The proof for this for near horizon geometries \cite{iiahor} is based on a partial integration argument instead.

\subsection{Counting of supersymmetries}

The counting of supersymmetries for $\mathrm{AdS}_2\times_w M^8$ backgrounds under the assumptions made in the previous section
is a special case of the proof of \cite{iiahor} that IIA horizons always preserve an even number of supersymmetries.  Here, we shall briefly
repeat the argument. If $N_\pm=\mathrm{dim}\, \mathrm{Ker}\, (\nabla_i^{(\pm)}, {\cal A}^{(\pm)})$, then the number of supersymmetries
preserved by the background is  $N=N_++N_-$. On the other hand from the Lichnerowicz type theorems of the previous section
\bea
N_\pm=\mathrm{dim}\, \mathrm{Ker}\, {\mathscr D}^{(\pm)}~.
\eea
Furthermore, it turns out that $\big(e^{2 \Phi} \Gamma_-\big) \big({\mathscr D}^{(+)}\big)^\dagger
= {\mathscr D}^{(-)} \big(e^{2 \Phi} \Gamma_-\big)$ and so
\bea
N_-=\mathrm{dim}\, \mathrm{Ker}\, {\mathscr D}^{(+)}{}^\dagger~.
\eea
On the other hand the index of ${\mathscr D}^{(+)}$ is the same as the index of the Dirac operator  $\slashed{\nabla}$ acting on the Majorana
representation of $Spin(8)$. The latter vanishes and so $N_+=N_-$.  Thus we conclude that $\mathrm{AdS}_2\times_w M^8$ solutions preserve
\bea
N=N_++N_-=2 N_-~,
\eea
supersymmetries confirming (\ref{cadssusy}).

\section{AdS$_3 \times_w {M}^7 $}

\subsection{Fields, Bianchi identities and field equations}

The fields of $\mathrm{AdS}_3$ backgrounds which are compatible with the $\mathrm{AdS}_3$ symmetries are
\bea
ds^2&=& 2 \bbe^+ \bbe^-+ A^2 dz^2+ds^2(M^7)~,
\cr
 G &=& A \bbe^+ \wedge \bbe^- \wedge dz \wedge X + Y~,~~~F=F~,
 \cr
 H &=& A W \bbe^+ \wedge \bbe^- \wedge dz + Z~,~~~S=S~,~~~\Phi=\Phi~,
 \label{fads3}
\eea
where
\bea
\bbe^+&=&du~,~~~\bbe^-=(dr+ rh)~,~~~\Delta=0
\cr
h&=&-{2\over\ell} dz-2 A^{-1} dA~,
\eea
$A$ is the warp factor which depends only on the coordinates of $M^7$, $(r,u,z)$ are the coordinates of $\mathrm{AdS}_3$, $X$ is a 1-form, $S, \Phi, W$ are functions, $F$ is a 2-form, $Z$ is a 3-form and $Y$
is a 4-form on $M^7$, respectively.

The Bianchi identities of (massive) IIA supergravity can now be rewritten as differential relations of the fields on $M^7$ as
\bea
 dZ &=& 0~,~~~
 d(A^3 W) = 0~,~~~
 dS = S d\Phi~,
 \cr
 dF &=& d\Phi \wedge F + S Z + A S W e^+ \wedge e^- \wedge dz~,~~~
 dY = d\Phi \wedge Y + Z \wedge F~,
 \cr
 dX &=& -3 A^{-1} dA \wedge X + d\Phi \wedge X - W F .
 \label{eq:f_bianchi}
\eea
The Bianchi identity involving $dF$ is consistent if either   $S=0$, or $W=0$. Therefore there are two distinct $\mathrm{AdS}_3$ backgrounds to consider.
One is a standard IIA supergravity background with a non-vanishing  component for $H$ on $\mathrm{AdS}_3$  or a massive IIA supergravity background with
$H$ that has components only along $M^7$.

Decomposing the field equations of (massive) IIA supergravity for the fields  (\ref{fads3}), one finds that
\bea
 \nabla^2 \Phi &=& -3 A^{-1} \partial_i A \partial^i \Phi  + 2 (d\Phi)^2 - \frac{1}{12} Z^2 + \frac{1}{2} W^2 + \frac{5}{4} S^2 + \frac{3}{8} F^2 + \frac{1}{96} Y^2 - \frac{1}{4} X^2~,
 \cr
 \nabla^k H_{i j k} &=& -3 A^{-1} \partial^k A H_{i j k} + 2 \partial^k \Phi H_{i j k} + \frac{1}{2} Y_{i j k \ell} F^{k \ell} + S F_{i j}~,
 \cr
 \nabla^j F_{i j} &=& -3 A^{-1} \partial^j A F_{i j} + \partial^j \Phi F_{i j} - W X_i - \frac{1}{6} Y_{i j k \ell} Z^{j k \ell}~,
 \cr
 \nabla^i X_i &=& \partial_i \Phi X^i - *_7 (Z \wedge Y)~,
 \cr
 \nabla^\ell Y_{i j k \ell} &=& -3 A^{-1} \partial^\ell A Y_{i j k \ell} + \partial^\ell \Phi Y_{i j k \ell} + *_7 (Z \wedge X - W Y) _{i j k}~,
\eea
and that the Einstein equation separates into an AdS component,
\begin{equation}
 \nabla^2 \ln A = -\frac{2}{\ell^2} A^{-2} - \frac{3}{\ell^2} A^{-2} (dA)^2 + 2 A^{-1} \partial_i A \partial^i \Phi + \frac{1}{2} W^2 + \frac{1}{4} S^2 + \frac{1}{8} F^2 + \frac{1}{96} Y^2 + \frac{1}{4} X^2
\end{equation}
and a transverse component,
\begin{align}
 R^{(7)}_{i j} &= 3 \nabla_i \nabla_j \ln A + 3 A^{-2} \partial_i A \partial_j A + \frac{1}{12} Y^2_{i j} - \frac{1}{2} X_i X_j - \frac{1}{96} Y^2 \delta_{i j}
 \\ \nonumber
 & \bigindent + \frac{1}{4} X^2 \delta_{i j} - \frac{1}{4} S^2 \delta_{i j} + \frac{1}{4} Z^2_{i j} + \frac{1}{2} F^2_{i j} - \frac{1}{8} F^2 \delta_{i j} - 2 \nabla_i \nabla_j \Phi~,
\end{align}
where $\nabla$ and  $R^{(7)}_{i j}$ are the Levi-Civita connection and the Ricci tensor of $M^7$, respectively.
The latter contracts to
\bea
 R^{(7)} &=& 3 \nabla^2 \ln A + 3 A^{-2} (dA)^2 + \frac{1}{4} Z^2 - \frac{7}{4} S^2 - \frac{3}{8} F^2 + \frac{1}{96} Y^2 + \frac{5}{4} X^2 - 2 \nabla^2 \Phi
 \cr
 &=& -\frac{6}{\ell^2} A^{-2} - 6 A^{-2} (dA)^2 + 12 A^{-1} \partial_i A \partial^i \Phi - 4 (d\Phi)^2 + \frac{5}{12} Z^2 + \frac{1}{2} W^2
 \cr
 && \bigindent - \frac{7}{2} S^2 - \frac{3}{4} F^2 + \frac{1}{48} Y^2 + \frac{5}{2} X^2~.
\eea
This form of the Ricci scalar is essential to establish the maximum principle formulae necessary for identifying the Killing spinors with the
zero modes of  Dirac-like operators.

\subsection{Local aspects: solution of KSEs}
\subsubsection{Solution of KSEs along $\mathrm{AdS}_3$}

The gravitino KSE along the $\mathrm{AdS}_3$ directions gives
\bea
 \partial_u \epsilon_\pm + A^{-1} \Gamma_{+ z} \left( \ell^{-1} - \Xi_- \right) \epsilon_\mp&=&0
 \cr
 \partial_r \epsilon_\pm - A^{-1} \Gamma_{- z} \Xi_+ \epsilon_\mp&=&0
 \cr
 \partial_z \epsilon_\pm - \Xi_\pm \epsilon_\pm + 2r\ell^{-1} A^{-1} \Gamma_{- z} \Xi_+ \epsilon_\mp&=&0
 \label{kseads33}
\eea
where
\begin{equation}
 \Xi_\pm = \mp \frac{1}{2 \ell} + \frac{1}{2} \slashed{\partial} A \Gamma_z \pm \frac{1}{4} A W \Gamma_{11} - \frac{1}{8} A S \Gamma_z - \frac{1}{16} A \slashed{F} \Gamma_z \Gamma_{11} - \frac{1}{192} A \slashed{Y} \Gamma_z \mp \frac{1}{8} A \slashed{X} .
\end{equation}
As in the $\mathrm{AdS}_2$ case, we integrate these equations along $r$ and $u$, and then along $z$. First observe that
\bea
\Theta_+=A^{-1} \Gamma_z \Xi_+~,~~~\Theta_-=A^{-1} \Gamma_z (\Xi_--\ell^{-1})~,
\eea
and that
\begin{align}
 \Xi_\pm \Gamma_{z +} + \Gamma_{z +} \Xi_\mp &= 0~,
 \\
 \Xi_\pm \Gamma_{z -} + \Gamma_{z -} \Xi_\mp &= 0~.
\end{align}
Integrating along the $r$ and $u$ coordinates, one finds that the Killing spinor can be expressed as in (\ref{ksesolads2}). To integrate along $z$ first note that
 the only  AdS-AdS integrability condition is
\begin{equation}
 \left( {\Xi_\pm}^2 \pm \ell^{-1} \Xi_\pm \right) \epsilon_\pm = 0 .
\end{equation}
Using this, one finds that the integration along  $z$ yields
\begin{equation}
 \eta_\pm  = \sigma_\pm  + e^{\mp z / \ell} \tau_\pm ~,
\end{equation}
where
\begin{equation}
 \Xi_\pm \sigma_\pm = 0 \qquad \Xi_\pm \tau_\pm = \mp \ell^{-1} \tau_\pm~,
 \label{addconads3}
\end{equation}
and  $\sigma_\pm, \tau_\pm$ are 16-component spinors counted over the reals, $\Gamma_\pm\sigma_\pm=\Gamma_\pm\tau_\pm=0$, that depend only on the coordinates of  $M^7$.

Combining all the above results together, one finds that the solution of the KSEs along $\mathrm{AdS}_3$ can be written as
\bea
\epsilon&=&\epsilon_++\epsilon_-
=\sigma_++ e^{-{z\over\ell}} \tau_++\sigma_-+e^{{z\over\ell}}\tau_-
\cr
&&-\ell^{-1} u A^{-1} \Gamma_{+z} \sigma_--\ell^{-1} r A^{-1}e^{-{z\over\ell}} \Gamma_{-z}\tau_+~,
\eea
where the dependence of $\epsilon$ on the $\mathrm{AdS}_3$ coordinates $(u,r,z)$ is given explicitly while the dependence on the coordinates $y$
of $M^7$ is via that of $\sigma_\pm, \tau_\pm$ spinors.

\subsubsection{Remaining independent KSEs}

As we have seen the KSEs of (massive) IIA supergravity have been solved provided that one imposes the additional conditions (\ref{addconads3}).
It is convenient to interpret these as new additional KSEs on $M^7$. In order to describe simultaneously the conditions on  both the $\sigma_\pm$ and $\tau_\pm$ spinors, we write $ \chi_\pm = \sigma_\pm, \tau_\pm $
and introduce
\bea
 \mathbb{B}^{(\pm)} &=& \mp \frac{c}{2 \ell} + \frac{1}{2} \slashed{\partial} A \Gamma_z \pm \frac{1}{4} A W \Gamma_{11} - \frac{1}{8} A S \Gamma_z
 \cr
 &&- \frac{1}{16} A \slashed{F} \Gamma_z \Gamma_{11} - \frac{1}{192} A \slashed{Y} \Gamma_z \mp \frac{1}{8} A \slashed{X} ,
\eea
where $c=1$ when $\chi_\pm=\sigma_\pm$ and $c=-1$ when $\chi_\pm=\tau_\pm$.

Using this, the remaining independent KSEs are
\bea
\nabla^{(\pm)}_i\chi_\pm=0~,~~~~\mathcal{A}^{(\pm)} \chi_\pm=0~,~~~ \mathbb{B}^{(\pm)}\chi_\pm=0~,
\label{indkseads3}
\eea
where
\bea
 \nabla_i^{(\pm)}&=& \nabla_i  + \Psi^{(\pm )}_i~,
 \cr
 \mathcal{A}^{(\pm)} &=& \slashed{\partial} \Phi + \frac{1}{12} \slashed{Z} \Gamma_{11} \mp \frac{1}{2} W \Gamma_z \Gamma_{11}
 \cr
 &&+ \frac{5}{4} S + \frac{3}{8} \slashed{F} \Gamma_{11} + \frac{1}{96} \slashed{Y} \pm \frac{1}{4} \slashed{X} \Gamma_z~,
\eea
and where
\bea
 \Psi^{( \pm )}_i = \pm \frac{1}{2} A^{-1} \partial_i A + \frac{1}{8} \slashed{Z}_i \Gamma_{11} + \frac{1}{8} S \Gamma_i + \frac{1}{16} \slashed{F} \Gamma_i \Gamma_{11} + \frac{1}{192} \slashed{Y} \Gamma_i \pm \frac{1}{8} \slashed{X} \Gamma_{z i}~.
\eea
It is clear that the first two equations in (\ref{indkseads3}) are the restrictions imposed on $\chi_\pm$ from gravitino and dilatino KSEs of (massive) IIA supergravity on $M^7$, while the last equation
has arisen from the integration of the supergravity KSEs on $\mathrm{AdS}_3$. All the other integrability conditions that arise in the analysis follow from (\ref{indkseads3}), the Bianchi
identities and the field equations.

\subsubsection{Counting supersymmetries}

The number of supersymmetries preserved by $\mathrm{AdS}_3\times_w M^7$ backgrounds  is the number of solutions of the KSEs (\ref{indkseads3}).
Thus
\bea
N=N_++N_-=(N_{\sigma_+}+ N_{\tau_+})+ (N_{\sigma_-}+ N_{\tau_-})~,
\eea
where $N_{\sigma_\pm}$ and $N_{\tau_\pm}$ denote the number of $\sigma_\pm$ and $\tau_\pm$ Killing spinors, respectively.
To prove that $\mathrm{AdS}_3$ backgrounds preserve an even number of supersymmetries as stated in (\ref{cadssusy}) observe that if $\chi_-$, for $\chi_-=\sigma_-$ or $\chi_-=\tau_-$, is a Killing spinor, ie
it solves all the three equations in (\ref{indkseads3}), then
\bea
\chi_+=A^{-1} \Gamma_{+z} \chi_-~,
\eea
also solves the KSEs (\ref{indkseads3}). Vice versa if $\chi_+$ solves the KSEs in (\ref{indkseads3}), then
\bea
\chi_-= A \Gamma_{-z} \chi_+~,
\label{chichiads3}
\eea
also solves the KSEs. Therefore $N_+=N_-$ and so $N=2 N_-$ which establishes (\ref{cadssusy}). Observe also that if $N_{\sigma_+}, N_{\tau_+}\not=0$ or  $N_{\sigma_-}, N_{\tau_-}\not=0$,
then $N=2 ( N_{\sigma_-}+N_{\tau_-})$ which refines (\ref{cadssusy}).

\subsection{Global aspects}

Here we shall demonstrate that the Killing spinors can be identified with the zero modes of a suitable Dirac-like operator on $M^7$.
We shall demonstrate this using the Hopf maximum principle as for the case of $\mathrm{AdS}_2\times_w M^8$ backgrounds.
As we have already mentioned the Bianchi identity for $F$ in (\ref{eq:f_bianchi}) implies that there are two different $\mathrm{AdS}_3\times_w M^7$ backgrounds to consider depending on whether the mass term vanishes and  $H$  is allowed to have a component along $\mathrm{AdS}_3$,
or the mass term does not vanish and $H$ has components only along $M^7$. Unlike the local analysis we have presented so far, the proof below of the Lichnerowicz type theorems
is sensitive to the two different cases and they will be investigated separately in appendix \ref{ads3co}.  However, the end result is the same including  coefficients in some key formulae.
Because of this and to save space, we shall present them together in the summary of the proof described below.

Furthermore, an argument similar to the one we have presented for $\mathrm{AdS}_2$ backgrounds implies that for smooth solutions $A$ does not vanish at any point on $M^7$.
This is based on the investigation of the field equation for $A$.

\subsubsection{Lichnerowicz type theorems for $ \sigma_\pm $ and $ \tau_\pm $}

To begin let us introduce the modified parallel transport operator
\bea
\hat{\nabla}^{(+)}_{i} = \nabla^{( + )}_{i} - {1\over7} A^{-1} \Gamma_{iz} \mathbb{B}^{( + )} -{1\over7} \Gamma_i \mathcal{A}^{(+)} ,
\eea
and the associated Dirac-like operator
\bea
{\mathscr D}^{(+)}=\slashed{\nabla}^{(+)}-A^{-1} \Gamma_{z} \mathbb{B}^{( + )}-\mathcal{A}^{(+)}~.
\eea

It is clear that if $\chi_+$ is a Killing spinor, for $\chi_+=\sigma_+$ or $\chi_+=\tau_+$, ie  satisfies the conditions (\ref{indkseads3}), then ${\mathscr D}^{(+)}\chi_+=0$.
To prove the converse suppose that ${\mathscr D}^{(+)}\chi_+=0$, then after some computation which utilizes the field equations, Bianchi identities (and has been presented
in appendix \ref{ads3co}),  one can establish the identity
\bea
\nabla^2\parallel \chi_+\parallel^2&+&(3 A^{-1}\partial_i A-2\partial_i\Phi) \nabla^i\parallel \chi_+\parallel^2= \parallel \hat\nabla^{(+)}\chi_+\parallel^2
+{16\over 7} \parallel A^{-1} \Gamma_z \mathbb{B}^{( + )}\chi_+\parallel^2
\cr
&&
+{4\over 7}\langle A^{-1} \Gamma_z \mathbb{B}^{( + )}\chi_+,\mathcal{A}^{(+)} \chi_+\rangle
+{2\over7} \parallel \mathcal{A}^{(+)} \chi_+\parallel^2~.
\label{maxads3}
\eea
First observe that the right-hand-side of the above expression is positive semi-definite. Applying the maximum principle on $\parallel \chi_+\parallel^2$, one concludes that
$\nabla^{(+)}\chi_+=\mathbb{B}^{( + )}\chi_+=\mathcal{A}^{(+)} \chi_+=0$ and that
\bea
\parallel \chi_+\parallel=\mathrm{const.}
\eea
Therefore $\chi_+$ is a Killing spinor.  Thus provided that the fields and $M^7$ satisfy the conditions for the maximum principle to apply,  we have established
that
\bea
\nabla^{(+)}_i\chi_+=0~,~~~\mathbb{B}^{( + )}\chi_+=0~,~~~\mathcal{A}^{(+)} \chi_+=0 \Longleftrightarrow {\mathscr D}^{(+)}\chi_+=0~.
\eea
It is remarkable that the zero modes of ${\mathscr D}^{(+)}$ satisfy all three KSEs.

Although we have presented Lichnerowicz type theorems for $ \sigma_+ $ and $ \tau_+ $ spinors, there is another similar theorem for
$ \sigma_- $ and $ \tau_- $ spinors. This can be established either  by a direct computation or by using (\ref{chichiads3}) which
relates the $\chi_+$ with the $\chi_-$ spinors. For this observe that in addition to the KSEs, the Clifford algebra operation $A\Gamma_{-z}$
intertwines between the corresponding Dirac-like operators ${\mathscr D}^{(+)}$ and ${\mathscr D}^{(-)}$.

\subsubsection{Counting supersymmetries again}

A consequence of the theorems of the previous section is that the number of supersymmetries of $\mathrm{AdS}_3\times_w M^7$ backgrounds
can be counted in terms of the zero modes of the Dirac-like operators ${\mathscr D}^{(\pm)}$.  In particular, one has
that
\bea
N=2 \big(\mathrm{dim}\, \mathrm{Ker}\, {\mathscr D}^{(-)}\vert_{c=1}+ \mathrm{dim}\, \mathrm{Ker}\, {\mathscr D}^{(-)}\vert_{c=-1}\big)~.
\eea
It is likely  that the dimension of these kernels, as the dimension of the Kernel of the standard Dirac operator,  depend on the geometry of $M^7$, ie they are not topological.

\section{AdS$_4 \times_w \mathcal{M}^6 $}

\subsection{Fields, Bianchi identities and field equations}

The fields of $\mathrm{AdS}_4\times_w M^6$ backgrounds are
\bea
ds^2 &=& 2 \bbe^+\bbe^-+ A^2 (dz^2+ e^{2z/\ell} dx^2) +ds^2(M^6)~,
\cr
G &= &A^2 e^{z/\ell} e^+ \wedge e^- \wedge dz \wedge dx X + Y~,~~~
\cr
H&=&H~,~~~F=F~,~~~\Phi=\Phi~,~~~S=S~,
\eea
where  $A, X, \Phi$ and $S$ are functions, $Y$ is a 4-form, $H$ is a 3-form and $F$ is a 2-form on $M^6$, respectively, and
\begin{eqnarray}
\bbe^+=du~,~~~\bbe^-=dr + r h~,~~~
h = -{2 \over \ell}dz -2 A^{-1} dA,~~~\Delta=0~.
\end{eqnarray}
$A$ is the warp factor.
The dependence of the fields  on the $\mathrm{AdS}_4$  coordinates  $(u,r,z,x)$ is given explicitly, while the dependence of the fields of
the coordinates $y$ of $M^6$ is suppressed.

The Bianchi identities of (massive) IIA supergravity impose the following conditions on the various components of the fields.

\bea
 dH &=& 0~,~~~
 dS = S d\Phi~,~~~
 dF = d\Phi \wedge F + S H~,
 \cr
 dY &=& d\Phi \wedge Y + H \wedge F~,~~~
 d(A^4 X) = A^4 d\Phi~.
\eea
Similarly, the field equations of the fluxes of (massive) IIA supergravity give
\bea
 \nabla^2 \Phi &=& -4 A^{-1} \partial^i A \partial_i \Phi + 2 ( d\Phi )^2 + \frac{5}{4} S^2 + \frac{3}{8} F^2 - \frac{1}{12} H^2 + \frac{1}{96} Y^2 - \frac{1}{4} X^2~,
 \cr
 \nabla^k H_{i j k} &=& -4 A^{-1} \partial^k A H_{i j k} + 2 \partial^k \Phi H_{i j k} + S F_{i j} + \frac{1}{2} F^{k \ell} G_{i j k \ell}~,
 \cr
 \nabla^j F_{i j} &=& -4 A^{-1} \partial^j A F_{i j} + \partial^j \Phi F_{i j} - \frac{1}{6} F^{j k \ell} G_{i j k \ell}~,
 \cr
 \nabla^\ell Y_{i j k \ell} &=& -4 A^{-1} \partial^\ell A Y_{i j k \ell} + \partial^\ell \Phi Y_{i j k \ell}~,
\eea
and the Einstein equation separates into an AdS component,
\bea
 \nabla^2 \ln A &=& -3 \ell^{-2} A^{-2} - 4 A^{-2} ( dA )^2 + 2 A^{-1} \partial_i A \partial^i \Phi + \frac{1}{96} Y^2
 \cr
 &&~~~~+ \frac{1}{4} X^2 + \frac{1}{4} S^2 + \frac{1}{8} F^2 ,
\eea
and a  component on $M^6$,
\begin{align} \nonumber
 R^{(6)}_{i j} &= 4 \nabla_i \nabla_j \ln A + 4 A^{-2} \partial_i A \partial_j A + \frac{1}{12} Y^2_{i j} - \frac{1}{96} Y^2 \delta_{i j} + \frac{1}{4} X^2 \delta_{i j}
 \\
 & \bigindent - \frac{1}{4} S^2 \delta_{i j} + \frac{1}{4} H^2_{i j} + \frac{1}{2} F^2_{i j} - \frac{1}{8} F^2 \delta_{i j} - 2 \nabla_i \nabla_j \Phi~,
\end{align}
where $R^{(6)}_{i j} $ is the Ricci tensor of $M^6$.
The latter contracts to
\bea
 R^{(6)} &=& 4 \nabla^2 \ln A + 4 A^{-2} ( dA )^2 + \frac{1}{48} Y^2 + \frac{3}{2} X^2 - \frac{3}{2} S^2 + \frac{1}{4} H^2 - \frac{1}{4} F^2 - 2 \nabla^2 \Phi
 \cr
 &=& - 12 \ell^{-2} A^{-2} - 12 A^{-2} \left( dA \right) ^2 + \frac{1}{24} Y^2 + 3 X^2 - 3 S^2 + \frac{5}{12} H^2
 \cr
 && \bigindent - \frac{1}{2} F^2 + 16 A^{-1} \partial_i A \partial^i \Phi - 4 ( d \Phi ) ^2 .
 \label{feqrads4}
\eea
This expression for the Ricci scalar is used in the proof of the Lichnerowicz type theorems for these backgrounds.

\subsection{Local aspects: Solution of KSEs}

\subsubsection{Solution of KSEs on $\mathrm{AdS}_4$}

The KSEs of (massive) IIA supergravity along the $\mathrm{AdS}_4$ directions give
\bea
\partial_u \epsilon_\pm + A^{-1} \Gamma_{+ z} \left( \ell^{-1} - \Xi_- \right) \epsilon_\mp&=&0~,
 \cr
 \partial_r \epsilon_\pm - A^{-1} \Gamma_{- z} \Xi_+ \epsilon_\mp&=&0~,
 \cr
\partial_z \epsilon_\pm - \Xi_\pm \epsilon_\pm + 2 r \ell^{-1} A^{-1} \Gamma_{- z} \Xi_+ \epsilon_\mp&=&0~,
 \cr
\partial_x \epsilon_+ + e^{z/\ell} \Gamma_{z x} \Xi_+ \epsilon_+&=&0~,
 \cr
 \partial_x \epsilon_- + e^{z/\ell} \Gamma_{z x} \left( \Xi_- - \ell^{-1} \right) \epsilon_-&=&0~,
 \label{kseads44}
\eea
where
\begin{equation}
 \Xi_\pm = \mp \frac{1}{2 \ell} + \frac{1}{2} \slashed{\partial} A \Gamma_z - \frac{1}{8} A S \Gamma_z - \frac{1}{16} A \slashed{F} \Gamma_z \Gamma_{11} - \frac{1}{192} A \slashed{Y} \Gamma_z \mp \frac{1}{8} A X \Gamma_x .
\end{equation}

Using
\bea
 \Xi_\pm \Gamma_{z +} + \Gamma_{z +} \Xi_\mp &=& 0~,~~~
 \Xi_\pm \Gamma_{z -} + \Gamma_{z -} \Xi_\mp = 0~,
 \cr
 \Xi_\pm \Gamma_{z x} + \Gamma_{z x} \Xi_\pm &=& \mp \ell^{-1} \Gamma_{z x},
\eea
one finds that there is only one integrability condition along all $\mathrm{AdS}_4$ directions,
\begin{equation}
 \left( {\Xi_\pm}^2 \pm \ell^{-1} \Xi_\pm \right) \epsilon_\pm = 0~.
\end{equation}
Thus, we can easily integrate the KSEs along $\mathrm{AdS}_4$. In particular, the integration along $r,u$ and $z$ proceeds as
for the $\mathrm{AdS}_3$ backgrounds.  Then integrating along $x$, we find that the Killing spinors can be expressed as

\bea
\epsilon&=&\epsilon_++\epsilon_-
=\sigma_+-\ell^{-1} x\Gamma_{xz} \tau_++ e^{-{z\over\ell}} \tau_++\sigma_-+e^{{z\over\ell}}(\tau_--\ell^{-1} x\Gamma_{xz} \sigma_-)
\cr
&&-\ell^{-1} u A^{-1} \Gamma_{+z} \sigma_--\ell^{-1} r A^{-1}e^{-{z\over\ell}} \Gamma_{-z}\tau_+~,
\eea
where
\begin{equation}
 \Xi_\pm \sigma_\pm = 0 \qquad \Xi_\pm \tau_\pm = \mp \ell^{-1} \tau_\pm~,
 \label{xiads4}
\end{equation}
and $\sigma_\pm$ and $\tau_\pm$ depend only on the coordinates of $M^6$. Observe that $\sigma_\pm$ and $\tau_\pm$ are again 16-component spinors counted over the reals.

\subsubsection{Remaining independent KSEs}

Having integrated the KSEs of (massive) IIA supergravity along the $\mathrm{AdS}_4$, it remains to identify the remaining independent KSEs.
For this, let us collectively denote $(\sigma_\pm, \tau_\pm)$ with $\chi_\pm$. It is also convenient to view (\ref{xiads4}) as additional KSEs on $M^6$. Investigating the various
integrability conditions that arise, one finds that the remaining independent KSEs are
\bea
\nabla_i^{( \pm )}\chi_\pm=0~,~~~\mathcal{A}^{(\pm)}\chi_\pm=0~,~~~ \mathbb{B}^{(\pm)}\chi_\pm=0~,
\label{indkseads4}
\eea
where
\bea
 \nabla_i^{(\pm)}&=&\nabla_i+\Psi^{(\pm)}_i~,~~~
 \cr
 \mathcal{A}^{(\pm)} &=& \slashed{\partial} \Phi + \frac{1}{12} \slashed{H} \Gamma_{11} + \frac{5}{4} S + \frac{3}{8} \slashed{F} \Gamma_{11} + \frac{1}{96} \slashed{Y} \mp \frac{1}{4} X \Gamma_{z x} ~,
 \cr
 \mathbb{B}^{(\pm)} &=& \mp \frac{c}{2 \ell} + \frac{1}{2} \slashed{\partial} A \Gamma_z - \frac{1}{8} A S \Gamma_z - \frac{1}{16} A \slashed{F} \Gamma_z \Gamma_{11}
 \cr&&~~~~~~- \frac{1}{192} A \slashed{Y} \Gamma_z \mp \frac{1}{8} A X \Gamma_x~,
\eea
and where
\begin{equation}
 \Psi^{( \pm )}_i = \pm \frac{1}{2 A} \partial_i A + \frac{1}{8} \slashed{H}_i \Gamma_{11} + \frac{1}{8} S \Gamma_i + \frac{1}{16} \slashed{F} \Gamma_i \Gamma_{11} + \frac{1}{192} \slashed{Y} \Gamma_i \mp \frac{1}{8} X \Gamma_{z x i}~.
\end{equation}
The constant $c$ in $ \mathbb{B}^{( \pm )} $ is chosen such that $c=1$ for $\chi_\pm=\sigma_\pm$ and $c=-1$ for $\chi_\pm=\tau_\pm$. Clearly, the first two equations in
(\ref{indkseads4}) arise from the gravitino and dilatino KSEs of (massive) IIA supergravity as adapted on the spinors $\chi_\pm$, respectively. The last equation in (\ref{indkseads4})
implements (\ref{xiads4}) on the spinors.

\subsubsection{Counting of supersymmetries}
The number of Killing spinors of  $\mathrm{AdS}_4$ backgrounds is
\bea
N=N_++N_-=(N_{\sigma_+}+ N_{\tau_+})+ (N_{\sigma_-}+ N_{\tau_-})~,
\eea
where $N_{\sigma_\pm}$ and $N_{\tau_\pm}$ denote the number of $\sigma_\pm$ and $\tau_\pm$ Killing spinors, respectively.

As for  $\mathrm{AdS}_3$ backgrounds one can verify by a direct computation that if $\chi_-$ is a Killing spinor, ie  solves (\ref{indkseads4}), then $\chi_+=A^{-1} \Gamma_{+z} \chi_-$ is also a Killing spinor, and vice-versa
if $\chi_+$ is a Killing spinor, then $\chi_-=A\Gamma_{-z} \chi_+$ is also a Killing spinor.  Furthermore, one can also verify that if $\tau_\pm$ is a Killing spinor, then
\bea
\sigma_\pm=\Gamma_{xz} \tau_\pm~,
\eea
is also a Killing spinor, and vice versa if $\sigma_\pm$  is a Killing spinor, then
\bea
\tau_\pm=\Gamma_{xz}\sigma_\pm~,
\eea
is a Killing spinor. As a result of this analysis, $N_{\sigma_+}=N_{\tau_+}=N_{\sigma_-}= N_{\tau_-}$ and so
\bea
N=4 N_{\sigma_-}~,
\eea
verifying (\ref{cadssusy}).

\subsection{Global aspects}

As in all previous cases, one can demonstrate that if the fields are smooth, then $A$ does not vanish at any point of $M^6$.
The argument is similar to that presented in the previous two cases and so it will not be repeated here.

\subsubsection{Lichnerowicz type theorems for $ \sigma_\pm $ and $ \tau_\pm $}

The Killing spinors $\sigma_\pm$ and $\tau_\pm$ of $\mathrm{AdS}_4$ backgrounds can be identified with the zero modes of a Dirac-like operator on $M^6$.
To determine this Dirac-like operator first define
\bea
\hat\nabla_i^{(\pm)}=\nabla_i^{(\pm)}-{1\over3} A^{-1} \Gamma_{iz} \mathbb{B}^{(\pm)}-{1\over6}\Gamma_i \mathcal{A}^{(\pm)}~.
\eea
and  the associated Dirac-like operator
\bea
{\mathscr D}^{(\pm)}\equiv \hat{\slashed{\nabla}}^{(\pm)} =\slashed{\nabla}^{(\pm)}-2A^{-1} \Gamma_{z} \mathbb{B}^{(\pm)}-\mathcal{A}^{(\pm)}~.
\eea
Then one can establish that
\bea
\nabla_i^{(\pm)}\chi_\pm=0~,~~~\mathbb{B}^{(\pm)}\chi_\pm=0~,~~~\mathcal{A}^{(\pm)}\chi_\pm=0 \Longleftrightarrow {\mathscr D}^{(\pm)}\chi_\pm=0~.
\label{kzads4}
\eea
It is apparent that if $\chi_\pm=(\sigma_\pm, \tau_\pm)$ are Killing spinors, then they are zero modes of ${\mathscr D}^{(\pm)}$. The task is to
demonstrate the converse.   We shall do this first for $\chi_+$ spinors. In particular let us assume that ${\mathscr D}^{(+)}\chi_+=0$. Then after
some extensive Clifford algebra calculus which is presented in appendix \ref{ads4co} and after using the Bianchi identities and the field equations, like (\ref{feqrads4}),
one can show that
\bea
&&\nabla^2\parallel\chi_+\parallel^2+(4 A^{-1}\partial^iA-2\partial^i\Phi)\nabla_i \parallel\chi_+\parallel^2=
\parallel\hat\nabla^{(+)}\chi_+\parallel^2
\cr &&~~~~~~
+ {16\over3} \parallel A^{-1}\Gamma_z \mathbb{B}^{(+)}\chi_+\parallel^2
+ {4\over3} \langle A^{-1} \Gamma_z\mathbb{B}^{(+)}\chi_+,
\mathcal{A}^{(+)}\chi_+\rangle
\cr&&~~~~~~ +{1\over3} \parallel \mathcal{A}^{(+)}\chi_+\parallel^2~.
\label{maxads4}
\eea
First observe that the right-hand-side of the above expression is positive semi-definite. Assuming that $M^6$ and the fields satisfy the requirements
for the application of the maximum principle to apply, eg $M^6$ compact without boundary and fields smooth, one concludes that $\chi_+$ is a Killing spinor and
in addition
\bea
\parallel\chi_+\parallel=\mathrm{const}~.
\eea
This proves (\ref{kzads4}) for the $\chi_+$ spinors.

To prove (\ref{kzads4}) for the $\chi_-$ spinors, one can either perform a similar computation to that of the $\chi_+$ spinors or simply use the
relation $\chi_-=A\Gamma_{-z} \chi_+$ between $\chi_+$ and $\chi_-$ spinors and observe that the Clifford algebra operation $A\Gamma_{-z}$ intertwines
between the Killing spinor equations and the Dirac-like operators. In particular, the analogous maximum principle relation to (\ref{maxads4})
for $\chi_-$ spinors can be constructed from (\ref{maxads4}) by simply setting $\chi_+=A^{-1}\Gamma_{+z} \chi_-$.

\subsubsection{Counting supersymmetries again}

A consequence of the theorems of the previous section is that one can count the  number of supersymmetries of $\mathrm{AdS}_4\times_w M^6$ backgrounds
in terms of the dimension of the Kernel of  ${\mathscr D}^{(\pm)}$ operators.  In particular, one has
that
\bea
N=4\, \mathrm{dim}\, \mathrm{Ker}\, {\mathscr D}^{(-)}\vert_{c=1}~.
\eea
As $\mathrm{dim}\,\mathrm{Ker}\, {\mathscr D}^{(-)}\vert_{c=1}=\mathrm{dim}\, \mathrm{Ker}\, {\mathscr D}^{(-)}\vert_{c=-1}=\mathrm{dim}\,\mathrm{Ker}\, {\mathscr D}^{(+)}\vert_{c=1}=\mathrm{dim}\,\mathrm{Ker}\, {\mathscr D}^{(+)}\vert_{c=-1}$, one can use  equivalently in the above formula the dimension of the Kernels of any of these operators.

\section{$\mathrm{AdS}_n \times_w {M}^{10 - n}$, $ n \geq 5 $}

\subsection{Fields, Bianchi identities and field equations}

For all $\mathrm{AdS}_n \times_w M^{10 - n} $, $ n \geq 5 $,  backgrounds, the form fluxes have non-vanishing components
only along $ M^{10 - n} $. In particular, the fields can be expressed as
\bea
ds^2 &=& 2 \bbe^+\bbe^-+ A^2 \big(dz^2+ e^{2z/\ell} \sum_{a=1}^{n-3}(dx^a)^2\big) +ds^2(M^{10-n})~,
\cr
G &= &G~,~~~
H=H~,~~~F=F~,~~~\Phi=\Phi~,~~~S=S~,
\eea
where  $A, \Phi$ and $S$ are functions, $G$ is a 4-form, $H$ is a 3-form and $F$ is a 2-form on $M^6$, respectively, and
\begin{eqnarray}
\bbe^+=du~,~~~\bbe^-=dr + r h~,~~~
h = -{2 \over \ell}dz -2 A^{-1} dA,~~~\Delta=0~.
\end{eqnarray}
$A$ is the warp factor and $\ell$ is the radius of $\mathrm{AdS}_n$.
The dependence of the fields  on the $\mathrm{AdS}_4$  coordinates  $(u,r,z,x^a)$ is given explicitly, while the dependence of the fields of
the coordinates $y$ of $M^{10-n}$ is suppressed. Clearly additional fluxes will vanish for large enough $n$, eg $\mathrm{AdS}_7$ backgrounds cannot have 4-form
fluxes, $G=0$.

 The Bianchi identities of the (massive) IIA supergravity give
\bea
 dH &=& 0~,~~~
 dS = S d\Phi~,~~~
 dF = d\Phi \wedge F + S H~,
 \cr
 dG &=& d\Phi \wedge G + H \wedge F~.
\eea
Furthermore, the field equations of (massive) IIA supergravity give
\begin{align}
 \nabla^2 \Phi &= -n A^{-1} \partial^i A \partial_i \Phi + 2 ( d\Phi )^2 + \frac{5}{4} S^2 + \frac{3}{8} F^2 - \frac{1}{12} H^2 + \frac{1}{96} G^2~,
 \\
 \nabla^k H_{i j k} &= -n A^{-1} \partial^k A H_{i j k} + 2 \partial^k \Phi H_{i j k} + S F_{i j} + \frac{1}{2} F^{k \ell} G_{i j k \ell}~,
 \\
 \nabla^j F_{i j} &= -n A^{-1} \partial^j A F_{i j} + \partial^j \Phi F_{i j} - \frac{1}{6} F^{j k \ell} G_{i j k \ell}~,
 \\
 \nabla^\ell G_{i j k \ell} &= -n A^{-1} \partial^\ell A G_{i j k \ell} + \partial^\ell \Phi G_{i j k \ell}~,
\end{align}
and the Einstein equation separates into an AdS component,
\begin{equation}
 \nabla^2 \ln A = -(n-1) \ell^{-2} A^{-2} - n A^{-2} ( dA )^2 + 2 A^{-1} \partial_i A \partial^i \Phi + \frac{1}{96} G^2 + \frac{1}{4} S^2 + \frac{1}{8} F^2 ,
\end{equation}
and $M^{10-n}$ component,
\begin{align}
 R^{(10-n)}_{i j} &= n \nabla_i \nabla_j \ln A + n A^{-2} \partial_i A \partial_j A + \frac{1}{12} G^2_{i j} - \frac{1}{96} G^2 \delta_{i j}
 \\ \nonumber
 & \bigindent - \frac{1}{4} S^2 \delta_{i j} + \frac{1}{4} H^2_{i j} + \frac{1}{2} F^2_{i j} - \frac{1}{8} F^2 \delta_{i j} - 2 \nabla_i \nabla_j \Phi~,
\end{align}
where $ R^{(10-n)}_{i j}$ is the Ricci tensor of $M^{10-n}$.
The latter contracts to
\bea
 R^{(10-n)} &=& n \nabla^2 \ln A + n A^{-2} ( dA )^2 + \frac{n - 2}{96} G^2 - \frac{10 - n}{4} S^2 + \frac{1}{4} H^2
 \cr&&~~~+\frac{n - 6}{8} F^2 - 2 \nabla^2 \Phi
 \cr
 &=&~~~- n ( n - 1 ) \ell^{-2} A^{-2} - n ( n - 1 ) A^{-2} \left( dA \right) ^2 + \frac{n - 2}{48} G^2
 \cr
 && - \frac{10-n}{2} S^2 + \frac{5}{12} H^2 + \frac{n-6}{4} F^2
 \cr&&~~~+ 4 n A^{-1} \partial_i A \partial^i \Phi - 4 ( d \Phi )^2~.
\eea
The expression for the Ricci scalar is essential for the proof of the Lichnerowicz type theorems below.

\subsection{Local aspects: Solution of KSEs}

\subsubsection{Solution of KSEs along $\mathrm{AdS}_n$}

The gravitino KSE of (massive) IIA supergravity along the $\mathrm{AdS}_n$ directions gives
\bea
\partial_u \epsilon_\pm + A^{-1} \Gamma_{+ z} \left( \ell^{-1} - \Xi_- \right) \epsilon_\mp&=&0~,
 \cr
 \partial_r \epsilon_\pm - A^{-1} \Gamma_{- z} \Xi_+ \epsilon_\mp&=&0~,
 \cr
\partial_z \epsilon_\pm - \Xi_\pm \epsilon_\pm + 2 r \ell^{-1} A^{-1} \Gamma_{- z} \Xi_+ \epsilon_\mp&=&0~,
 \cr
\partial_a \epsilon_+ + e^{z/\ell} \Gamma_{ za} \Xi_+ \epsilon_+&=&0~,
 \cr
 \partial_a \epsilon_- + e^{z/\ell} \Gamma_{za} \left( \Xi_- - \ell^{-1} \right) \epsilon_-&=&0~,
 \label{kseadsnn}
\eea
where
\begin{equation}
 \Xi_\pm = \mp \frac{1}{2 \ell} + \frac{1}{2} \slashed{\partial} A \Gamma_z - \frac{1}{8} A S \Gamma_z - \frac{1}{16} A \slashed{F} \Gamma_z \Gamma_{11} - \frac{1}{192} A \slashed{G} \Gamma_z~.
\end{equation}

Using the identities,
\begin{align}
 \Xi_\pm \Gamma_{z +} + \Gamma_{z +} \Xi_\mp &= 0~,
 \\
 \Xi_\pm \Gamma_{z -} + \Gamma_{z -} \Xi_\mp &= 0~,
 \\
 \Xi_\pm \Gamma_{z a} + \Gamma_{z a} \Xi_\pm &= \mp \ell^{-1} \Gamma_{z a},
\end{align}
ones finds that all these equations can be solved provided the integrability condition
\begin{equation}
 \left( {\Xi_\pm}^2 \pm \ell^{-1} \Xi_\pm \right) \epsilon_\pm = 0~,
\end{equation}
 is satisfied. In particular, one finds that the Killing spinor can be expressed as
 \bea
\epsilon&=&\epsilon_++\epsilon_-
=\sigma_+-\ell^{-1} \sum_{a=1}^{n-3} x^a\Gamma_{az} \tau_++ e^{-{z\over\ell}} \tau_++\sigma_-+e^{{z\over\ell}}(\tau_--\ell^{-1} \sum_{a=1}^{n-3} x^a \Gamma_{az} \sigma_-)
\cr
&&-\ell^{-1} u A^{-1} \Gamma_{+z} \sigma_--\ell^{-1} r A^{-1}e^{-{z\over\ell}} \Gamma_{-z}\tau_+~,
\eea
where
\begin{equation}
 \Xi_\pm \sigma_\pm = 0 \qquad \Xi_\pm \tau_\pm = \mp \ell^{-1} \tau_\pm~,
 \label{xiadsn}
\end{equation}
and $\sigma_\pm$ and $\tau_\pm$ are 16-component spinors depending only on the coordinates of $M^{10-n}$.
The dependence of the Killing spinors on the $\mathrm{AdS}_n$ coordinates is given explicitly while that of the coordinates $y$
of $M^{10-n}$ is via the $\sigma_\pm$ and $\tau_\pm$ spinors.

\subsubsection{Remaining independent KSEs}

Having solved the gravitino KSE along  $\mathrm{AdS}_n$,  $n>4$, to count the number of supersymmetries preserved by these backgrounds, one has to identify the remaining independent KSEs.
There are several integrability conditions which have to be considered.  However after using the field equations and the Bianchi identities, one finds that
the remaining independent KSEs are
\bea
 \nabla^{( \pm )}_i\chi_\pm=0~,~~~\mathcal{A}^{(\pm)}\chi_\pm=0~,~~~ \mathbb{B}^{(\pm)} \chi_\pm=0~,
 \label{indkseadsn}
\eea
where
\bea
\nabla^{( \pm )}_i&=& \nabla_i + \Psi^{( \pm )}_i~,
\cr
\mathcal{A}^{(\pm)} &=&\slashed{\partial} \Phi + \frac{1}{12} \slashed{H} \Gamma_{11} + \frac{5}{4} S + \frac{3}{8} \slashed{F} \Gamma_{11} + \frac{1}{96} \slashed{G}\big)
\cr
\mathbb{B}^{(\pm)} &=& \mp \frac{c}{2 \ell} + \frac{1}{2} \slashed{\partial} A \Gamma_z - \frac{1}{8} A S \Gamma_z - \frac{1}{16} A \slashed{F} \Gamma_z \Gamma_{11} -
\frac{1}{192} A \slashed{G} \Gamma_z~,
\eea
and where
\begin{equation}
 \Psi^{( \pm )}_i = \pm \frac{1}{2 A} \partial_i A + \frac{1}{8} \slashed{H}_i \Gamma_{11} + \frac{1}{8} S \Gamma_i + \frac{1}{16} \slashed{F} \Gamma_i \Gamma_{11} + \frac{1}{192} \slashed{G} \Gamma_i~.
\end{equation}
We have also set $\chi_\pm=(\sigma_\pm, \tau_\pm)$, and $c=1$ whenever $\chi_\pm=\sigma_\pm$ and $c=-1$ whenever $\chi_\pm=\tau_\pm$.

The first two KSEs in (\ref{indkseadsn}) arise from gravitino and dilatino KSEs of (massive) IIA supergravity as they are implemented on $\chi_\pm$, respectively. The last
equation in (\ref{indkseadsn}) is the condition (\ref{xiadsn}) which is now interpreted as additional algebraic KSE.
All the remaining integrability conditions are implied from (\ref{indkseadsn}), the Bianchi identities and the field equations.

\subsubsection{Counting supersymmetries}
\label{adsncount}

As in previous cases, the number of supersymmetries $N$ of $\mathrm{AdS}_n$ backgrounds is
\bea
N=N_++N_-=(N_{\sigma_+}+ N_{\tau_+})+ (N_{\sigma_-}+ N_{\tau_-})~,
\eea
where $N_{\sigma_\pm}$ and $N_{\tau_\pm}$ denote the number of $\sigma_\pm$ and $\tau_\pm$ Killing spinors, respectively.

 A direct inspection of the remaining independent KSEs (\ref{indkseadsn}) reveals that if $\chi_-$ is a solution,  then so is  $\chi_+=A^{-1} \Gamma_{+z} \chi_-$, and vice-versa
if $\chi_+$ is a Killing spinor, then $\chi_-=A\Gamma_{-z} \chi_+$ is also a Killing spinor.  Therefore $N_+=N_-$. Moreover to count the number of supersymmetries it suffices
to count the number of $\chi_-$ spinors.

 Furthermore if $\tau_-$ is a  Killing spinor, then $\sigma_-=\Gamma_{az} \tau_-$ is also a Killing spinor,  and vice versa if $\sigma_-$  is a Killing spinor, then
$\tau_-=\Gamma_{az}\sigma_-$
is a Killing spinor.  Thus $N_{\sigma_-}=N_{\tau_-}$ and so  $N=4 N_{\sigma_-}$. Therefore, it remains to count the number of $\sigma_-$ Killing spinors.

For this observe that
if  $\sigma_-$ is a Killing spinor, then
\bea
\sigma'_-=\Gamma_{ab} \sigma_-~,~~~a<b~,
\eea
is also a Killing spinor. To find  $N_{\sigma_-}$, one has to count the number of linearly independent $(\sigma, \Gamma_{ab} \sigma_-), a<b$ spinors.  This depends on $n$.
For $n=5$, $a,b=1,2$ and $(\sigma, \Gamma_{12}\sigma)$ are linearly independent. Thus $\mathrm{AdS}_5$ backgrounds preserve $N=8k$ supersymmetries.  Next for $n=6$, $a,b=1,2,3$
and $(\sigma, \Gamma_{12}\sigma, \Gamma_{13}\sigma, \Gamma_{23}\sigma)$ are linearly independent. Thus $\mathrm{AdS}_6$ backgrounds preserve $N=16k$ supersymmetries. To continue
for $n=7$, $a,b=1,2,3,4$. It turns out that in this case the Clifford algebra operation $\Gamma_{1234}$ commutes with all KSEs and therefore one can impose
consistently $\Gamma_{1234}\sigma_-^\pm=\pm \sigma_-^\pm$, ie one can restrict $\sigma_-$ to lie in one of the eigenspaces of $\Gamma_{1234}$. In such a case,
there are only 4 linearly independent spinors $(\sigma_-, \Gamma_{ab}\sigma_-)$, $a<b$.  Thus $\mathrm{AdS}_7$ backgrounds again preserve $16k$ supersymmetries. These results
confirm the counting of supersymmetries as stated in (\ref{cadssusy}).

There are no $\mathrm{AdS}_n$, $n>7$ backgrounds. This can be seen as follows. If the counting of supersymmetries proceeds in the same way one can show that all such backgrounds
preserve 32 supersymmetries.  The maximally supersymmetric backgrounds of (massive) IIA supergravity have been classified in \cite{maxsusy} and they do not include $\mathrm{AdS}_n\times_w M^{10-n}$
 spaces. The same result can be used to rule out the existence of $\mathrm{AdS}_7$ backgrounds that preserve 32 supersymmetries.

\subsection{Global aspects}

\subsubsection{Lichnerowicz type theorems for $ \sigma_\pm $ and $ \tau_\pm $}
As in all previous cases, the Killing spinors $\chi_\pm$ of the $\mathrm{AdS}_n$, $n>4$, backgrounds can be identified with the zero modes
of a suitable Dirac-like operator. To prove this first define
\bea
\hat{{\nabla}}_i^{(\pm)}={\nabla}^{(\pm)}_i-{n-2\over 10-n} A^{-1} \Gamma_{iz} \mathbb{B}^{(\pm)}-{1\over10-n}  \Gamma_i \mathcal{A}^{(\pm)}~,
\eea
and
\bea
{\mathscr D}^{(\pm)}\equiv \hat{\slashed{\nabla}}^{(\pm)} =\slashed{\nabla}^{(\pm)}-(n-2)A^{-1} \Gamma_{z} \mathbb{B}^{(\pm)}-  \mathcal{A}^{(\pm)}~.
\eea

Then one can show that
\bea
{\nabla}^{(\pm)}\chi_\pm=0~,~~~\mathcal{A}^{(\pm)}\chi_\pm=0~,~~~\mathbb{B}^{(\pm)}\chi_\pm=0 \Longleftrightarrow {\mathscr D}^{(\pm)}\chi_\pm=0~.
\eea
Clearly the proof of this statement in the forward direction is straightforward. The main task is to prove the converse.  It suffices to show this for $\chi_+$ spinors.  This is because the Clifford algebra operations $\chi_+=A^{-1} \Gamma_{+z} \chi_-$ and  $\chi_-=A\Gamma_{-z} \chi_+$
which relate these spinors intertwine between the corresponding KSEs and the Dirac-like operators.

Next suppose that $\chi_+$ is a zero mode of the ${\mathscr D}^{(+)}$ operator, ${\mathscr D}^{(+)}\chi_+=0$. Then after some computation which is presented in appendix \ref{adsnco}
which involves the use of the field equations and Bianchi identities, one finds that
\bea
 && \nabla^2 \left\| \chi_+ \right\| ^2 + \left( n A^{-1} \partial_i A - 2 \partial_i \Phi \right) \nabla^i \left\| \chi_+ \right\| ^2=\left\| \hat{\nabla} \chi_+ \right\| ^2
 \cr
 &&~~~
  + \frac{16 (n-2)}{10-n}  \left\| A^{-1} \Gamma_z \mathbb{B}^{(+)} \chi_+ \right\| ^2 + \frac{4 (n-2)}{10-n} \left\langle A^{-1} \Gamma_z \mathbb{B}^{(+)} \chi_+, \mathcal{A}^{(+)} \chi_+ \right\rangle
  \cr
 &&~~~~~~~~+ \frac{2}{10-n} \left\| \mathcal{A}^{(+)} \chi_+ \right\| ^2~.
 \label{maxadsn}
\eea
To proceed one has to solve the above differential equations. For this  observe that if the fields are smooth $A$ does not vanish at any point of $M^{10-n}$.
The proof of this is similar to that presented in the  previous cases. Furthermore, the right-hand-side of (\ref{maxadsn}) is positive semi-definite. Thus if $M^{10-n}$ and the fields
satisfy the conditions for the application of the maximum principle, eg $M^{10-n}$ compact without boundary and the fields smooth, then the only solution of this is
that $\chi_+$ is a Killing spinor and that
\bea
\left\| \chi_+ \right\|^2 =\mathrm{const}~.
\eea
This completes the proof of the theorem.

\subsubsection{Counting supersymmetries again}

A consequence of  the results of the previous section is that   the  number of supersymmetries of $\mathrm{AdS}_n\times_w M^{10-n}$ backgrounds
can be expressed in terms of the dimension of the Kernel of  ${\mathscr D}^{(\pm)}$ operators.  In particular, one has
that
\bea
N=4\, \mathrm{dim}\, \mathrm{Ker}\, {\mathscr D}^{(-)}\vert_{c=1}~.
\eea
 Equivalently, $N$ can be expressed in terms of $\mathrm{dim}\, \mathrm{Ker}\, {\mathscr D}^{(-)}\vert_{c=-1}$, $\mathrm{dim}\,\mathrm{Ker}\, {\mathscr D}^{(+)}\vert_{c=1}$ and $\mathrm{dim}\,\mathrm{Ker}\, {\mathscr D}^{(+)}\vert_{c=-1}$ as all these numbers are equal. Furthermore $\mathrm{dim}\,\mathrm{Ker}\, {\mathscr D}^{(-)}\vert_{c=1}$ has  multiplicity $2^{[{n\over2}]-1}$.  This can be seen
 by an analysis similar to that we have done for the counting the supersymmetries of these backgrounds in section \ref{adsncount}.

\begin{table}
\centering
\fontencoding{OML}\fontfamily{cmm}\fontseries{m}\fontshape{it}\selectfont
\begin{tabular}{|c|c|}\hline
$AdS_n\times_w M^{10-n}$& $N$
 \\
\hline\hline

$n=2$&$2k, k\leq15$
\\
\hline
$n=3$&$2k, k\leq15$
\\
\hline
$n=4$&$4k, k\leq7$
\\
\hline
$n=5$&$8k, k\leq 3$
\\
\hline
$n=6, 7$&$16$
\\
\hline
$n> 7$&$-$
\\
\hline
\end{tabular}
\label{tab2}
\begin{caption}
{\small {\rm ~~The number of supersymmetries $N$ of $AdS_n\times_w M^{10-n}$ backgrounds are given. For $AdS_2\times_w M^{8}$, one can show that these backgrounds preserve an even number of supersymmetries
provided that  $M^8$ and the fields satisfy the maximum principle. For the counting of supersymmetries of  the rest of the backgrounds such an assumption is not necessary. The bounds on $k$ arise from the non-existence
 of supersymmetric solutions with  maximal supersymmetry. For the remaining fractions, it is not known whether there always exist backgrounds preserving the prescribed
 number of supersymmetries. Supersymmetric $AdS_n$, $n> 7$, backgrounds do not exist.
}}
\end{caption}
\end{table}

\section{Flux $\bR^{n-1,1}\times_w M^{10-n}$ backgrounds}

In the limit of large $\mathrm{AdS}$ radius $\ell$, $\mathrm{AdS}_n\times_w M^{10-n}$  become warped flux $\bR^{n-1,1}\times_w M^{10-n}$ backgrounds.
Furthermore all the local computations we
have performed for $\mathrm{AdS}_n\times_w M^{10-n}$ backgrounds are still valid after taking $\ell\rightarrow \infty$ and so they can be used to investigate the $\bR^{n-1,1}\times_w M^{10-n}$ backgrounds.  These include the expressions for the fields, Bianchi identities,
field equations as well as  the local solutions to the KSEs, and the determination  of the independent KSEs on $M^{10-n}$.

However, there are some differences as well.  First the counting of supersymmetries is different.  This is because the criteria for the linear independence of the solutions
of the KSEs on $M^{10-n}$ for $\mathrm{AdS}_n\times_w M^{10-n}$ backgrounds are different from those of  $\bR^{n-1,1}\times_w M^{10-n}$ backgrounds.
Secondly,  the global properties of the KSEs  for  $\mathrm{AdS}_n\times_w M^{10-n}$ and $\bR^{n-1,1}\times_w M^{10-n}$ backgrounds are different, which originates in differences
between
the regularity properties of $\mathrm{AdS}_n\times_w M^{10-n}$ and $\bR^{n-1,1}\times_w M^{10-n}$ backgrounds. It is well known for example that there are
no smooth flux compactifications of supergravity theories to $\bR^{n-1,1}$ with a compact\footnote{We shall demonstrate below that the same conclusion applies under a weaker hypothesis.} internal space $M^{10-n}$.

\subsection{Non-existence of flux $\bR^{n-1,1}\times_w M^{10-n}$ backgrounds and maximum principle}

One of the main properties of $\mathrm{AdS}$ backgrounds is that the warp factor $A$ can be no-where vanishing even if  $M^{10-n}$ is compact.  This
is essential for the regularity.  As we have seen, this property relies on the  radius $\ell$ of $\mathrm{AdS}$ and it is no longer valid
in the limit $\ell\rightarrow \infty$.

In fact one can show that the only  $\bR^{n-1,1}\times_w M^{10-n}$ backgrounds of (massive)  IIA supergravity for which the fields and $M^{10-n}$ are chosen such that the  maximum principle applies are those for
which all fluxes vanish, and the dilaton and warp factor are constant. To see this, observe that the field equation of the warp factor $A$ in all cases
can be rewritten as a  differential inequality
\bea
\nabla^2 \ln A+ b^i \partial_i \ln A=\Sigma\geq 0~,
\eea
for some $b$ which depends on $A$ and the dilaton and $\Sigma$ which depends again on the fields. Therefore it is in a form that the maximum principle can apply. Assuming that the maximum principle applies,  the only solution of this equation  is that
 $A$ is constant and $\Sigma=0$. The latter condition in turn gives that all the fluxes must vanish apart from the component of $H$ on $M^{10-n}$ and the dilaton  which are not restricted. However the vanishing of the rest of the fields turns the field equation for the dilaton into a maximum principle
 form. Applying the maximum principle again for this, one finds that the dilaton is constant and the component of $H$ on $M^{10-n}$ vanishes as well.
Therefore there are no warped flux $\bR^{n-1,1}\times_w M^{10-n}$  backgrounds which satisfy the maximum principle. Observe that this result applies irrespective on whether
the solution is supersymmetric or not.

In the context of flux compactifications based on $\bR^{n-1,1}\times_w M^{10-n}$  this no-go theorem may be circumvented in various ways. One way is to take  $M^{10-n}$ to be non-compact.
Another way is to no longer assume that various fields
satisfy the properties required for the maximum principle to hold, by
weakening the assumption of smoothness.
One can also add brane charges which modify the Bianchi identities and the field equations,
and/or add higher order corrections.  However here we shall focus on the properties of supergravity and we shall simply assume that the fields and $M^{10-n}$ do not satisfy the requirements for maximum principle to apply.

\subsection{Supersymmetry of flux $\bR^{n-1,1}\times_w M^{10-n}$ backgrounds}

\subsubsection{$\bR^{1,1}\times_w M^{8}$}

The proof that  $\mathrm{AdS}_2\times_w M^8$ backgrounds preserve an even number of supersymmetries relies on the maximum principle which is not  applicable
to $\bR^{1,1}\times_w M^{8}$ supergravity backgrounds.  Because of this, we cannot establish in generality that flux $\bR^{1,1}\times_w M^{8}$ backgrounds
preserve an even number of supersymmetries.  Nevertheless some supersymmetry enhancement is expected. In particular, we have seen that it is a property of (massive) IIA supergravity that if $\eta_-$ is a Killing spinor
then $\eta_+=\Gamma_+\Theta_-\eta_-$ is also a Killing spinor. Supersymmetry enhancement takes place whenever $\eta_-\notin \mathrm{Ker}\, \Theta_-$ and so $\eta_+\not=0$.  However
there is no general argument which leads to $\eta_+\not=0$ and so this has to be established on a case by case basis.

The general form of the Killing spinor is
\bea
\epsilon=\eta_++\eta_-+ u\Gamma_+\Theta_-\eta_-+ r \Gamma_-\Theta_+ \eta_+~,
\eea
for a general choice of $\eta_\pm$.  To establish the above expression from that  in (\ref{ksesolads2}) for $\mathrm{AdS}_2$ backgrounds, we have taken the limit $\ell\rightarrow \infty$ and we have used the integrability conditions of the KSEs stated in \cite{iiahor} which  read
\bea
 \Gamma_\mp\Theta_\pm\Gamma_\pm\Theta_\mp\eta_\mp=0~.
 \eea
These are automatically satisfied as a consequence of the independent KSEs on $M^8$ (\ref{kseadsm8}), the Bianchi identities and the field equations.
Note that the Killing spinor $\epsilon$ is at most linear in the coordinates $(u,r)$ of $\bR^{1,1}$. This conclusion arises from the general analysis
we have done and it is contrary to the expectation that the Killing spinors of flux $\bR^{1,1}\times_w M^{8}$ backgrounds do not depend on the coordinates of  $\bR^{1,1}$.
Notice also that $\epsilon$ does not depend on $(u,r)$ whenever $\eta_\pm$ are in the Kernel of $\Theta_\pm$. We shall further comment on these below.

\subsubsection{$\bR^{2,1}\times_w M^{7}$}

The solution of the KSEs (\ref{kseads33}) in the limit $\ell\rightarrow \infty$  is
\bea
\epsilon=\sigma_++\sigma_-+ u\Gamma_{+z} \Xi_-\sigma_-+ r\Gamma_{-z} \Xi_+\sigma_++ z (\Xi_+\sigma_++\Xi_-\sigma_-)~,
\eea
provided that the integrability conditions
\bea
(\Xi_\pm)^2\sigma_\pm=0~,
\eea
are satisfied,
where $\sigma_\pm$ depend only on the coordinates of $M^7$. Moreover necessary and sufficient conditions for $\epsilon$ to be a Killing spinor are that  $\sigma_\pm$ must satisfy the KSEs (\ref{indkseads3}) on $M^7$.

Comparing the above result with that for $\mathrm{AdS}_3\times_w M^7$ backgrounds, one notices that the $\tau_\pm$ spinors do not arise.  This is because the $\tau_\pm$ spinors
are not linearly independent from the $\sigma_\pm$ ones for $\bR^{2,1}\times_w M^{7}$ backgrounds. The same applies for the rest of $\bR^{n-1,1}\times_w M^{10-n}$ backgrounds
and so the explanation will not be repeated below.

To count the number $N$ of supersymmetries preserved by the $\bR^{2,1}\times_w M^{7}$ backgrounds, first observe that $N=N_{\sigma_+}+N_{\sigma_-}$, where $N_{\sigma_+}$ and $N_{\sigma_-}$ is the number of $\sigma_+$ and $\sigma_-$ Killing spinors, respectively. Then notice that if $\sigma_-$ is a Killing spinor, then $\sigma_+=A^{-1}\Gamma_{+z} \sigma_-$
is also a Killing spinor, and vice versa if $\sigma_+$ is a Killing spinor then $\sigma_-= A \Gamma_{-z}\sigma_+$ is also a Killing spinor.  Therefore $N_{\sigma_+}=N_{\sigma_-}$, and so
$N=2 N_{\sigma_-}$, ie the $\bR^{2,1}\times_w M^{7}$ solutions preserve an even number of supersymmetries confirming (\ref{cafsusy}).

\subsubsection{$\bR^{3,1}\times_w M^{6}$}

The solution of the KSEs (\ref{kseads44}) in the limit $\ell\rightarrow \infty$ is
\bea
\epsilon=\sigma_++\sigma_-+ u\Gamma_{+z} \Xi_-\sigma_-+ r\Gamma_{-z} \Xi_+\sigma_++ (z+x\Gamma_{xz}) (\Xi_+\sigma_++\Xi_-\sigma_-)~,
\eea
provided that the integrability conditions
\bea
(\Xi_\pm)^2\sigma_\pm=0~,
\eea
are satisfied,
where $\sigma_\pm$ depend only on the coordinates of $M^6$. Moreover necessary and sufficient conditions for $\epsilon$ to be a Killing spinor are that  $\sigma_\pm$ must satisfy the KSEs (\ref{indkseads4}) on $M^6$.

The number of supersymmetries preserved by the $\bR^{3,1}\times_w M^{6}$ backgrounds is $N=N_{\sigma_+}+N_{\sigma_-}$ where $N_{\sigma_+}$ and $N_{\sigma_-}$ is the number of $\sigma_+$ and $\sigma_-$ Killing spinors, respectively. Furthermore as in the  $\bR^{2,1}\times_w M^{7}$ case above $N_{\sigma_+}=N_{\sigma_-}$.  In addition, if $\sigma_\pm$ is a Killing spinor
so is $\sigma_\pm'=\Gamma_{zx} \sigma_\pm$. As a result $N_{\sigma_\pm}$ are even numbers.  Thus $\bR^{3,1}\times_w M^{6}$ backgrounds preserve $4k$ supersymmetries
confirming  (\ref{cafsusy}).

\subsubsection{$\bR^{n-1,1}\times_w M^{10-n}$ for $n\geq 5$}

The solution of the KSEs (\ref{kseadsnn}) in the limit $\ell\rightarrow \infty$ is
\bea
\epsilon=\sigma_++\sigma_-+ u\Gamma_{+z} \Xi_-\sigma_-+ r\Gamma_{-z} \Xi_+\sigma_++ (z+\sum_{a=1}^{n-3}x^a\Gamma_{az}) (\Xi_+\sigma_++\Xi_-\sigma_-)~,
\eea
provided that the integrability conditions
\bea
(\Xi_\pm)^2\sigma_\pm=0~,
\eea
are satisfied,
where $\sigma_\pm$ depend only on the coordinates of $M^{10-n}$. Moreover necessary and sufficient conditions for $\epsilon$ to be a Killing spinor are that  $\sigma_\pm$ must satisfy the KSEs (\ref{indkseadsn}) on $M^{10-n}$.

To count the number of supersymmetries preserved by these backgrounds observe that $N=N_{\sigma_+}+N_{\sigma_-}$  and that $N_{\sigma_+}=N_{\sigma_-}$ as in  previous cases.
Therefore it suffices to count the multiplicity of $\sigma_-$ Killing spinors. For this notice that for $\bR^{n-1,1}\times_w M^{10-n}$ backgrounds, the $z$ coordinate
can be treated in the same way as the $x^a$ coordinates. As a result let us denote with $x^{a'}=(z, x^a)$ all the coordinates of $\bR^{n-1,1}$ transverse to the lightcone.
Furthermore observe that if $\sigma_-$ is a Killing spinor so is $\Gamma_{a'b'}\sigma_-$ for $a'<b'$.  Therefore it suffices to count the linearly independent
$(\sigma_-, \Gamma_{a'b'}\sigma_-)$, $a'<b'$ spinors in each case. For the analysis that follows, we shall choose directions for convenience and therefore the analysis  is not  fully covariant.  However, it can be made covariant as that presented in \cite{mads}.

For $\bR^{4,1}\times_w M^{5}$ a direct computation reveals that there are 4 linearly independent $(\sigma_-, \Gamma_{a'b'}\sigma_-)$, $a'<b'$, $a',b'=1,2, 3$, spinors
leading to the conclusion that such backgrounds preserve $N=8k$ supersymmetries.

For $\bR^{5,1}\times_w M^{4}$, one can impose the projection $\Gamma_{1234}\sigma^\pm_-=\pm \sigma^\pm$
as $a',b'=1,2,3,4$ and since $\Gamma_{1234}$ commutes with all KSEs. If $\sigma_-$ is chosen to be in one of the two eigenspaces of $\Gamma_{1234}$, then only 4 of the $(\sigma_-, \Gamma_{a'b'}\sigma_-)$, $a'<b'$, spinors are linearly independent.  As a result, $\bR^{5,1}\times_w M^{4}$ backgrounds preserve $N=8k$ supersymmetries as well.

A similar argument implies to the counting of supersymmetries for
$\bR^{6,1}\times_w M^{3}$ backgrounds. Imposing that $\sigma_-$ lies in one of the eigenspaces of $\Gamma_{1234}$, only 8 of the spinors  $(\sigma_-, \Gamma_{a'b'}\sigma_-)$, $a'<b'$,
$a', b'=1,2,3,4,5$ are linearly independent.  Therefore these backgrounds  preserve $16k$ supersymmetries.

For $\bR^{7,1}\times_w M^{2}$ backgrounds, $\sigma_-$ can be chosen to lie in an eigenspace
of two Clifford algebra operators, say $\Gamma_{1234}$ and $\Gamma_{1256}$.  In such a case only 8 of the spinors  $(\sigma_-, \Gamma_{a'b'}\sigma_-)$, $a'<b'$,
$a', b'=1,2,3,4,5, 6$ are linearly independent and so such backgrounds also preserve $16k$ supersymmetries.

Next consider the
 $\bR^{8,1}\times_w M^{1}$ backgrounds which include the D8-brane solution. In this case  $\sigma_-$ can be chosen to lie  in an eigenspace of  $\Gamma_{1234}$, $\Gamma_{1256}$ and $\Gamma_{1357}$. For such a choice,  there are  only 8 of the spinors  $(\sigma_-, \Gamma_{a'b'}\sigma_-)$, $a'<b'$, $a', b'=1,2,3,4,5, 6, 7$ are linearly independent. Therefore such backgrounds also
  preserve $N=16k$ supersymmetries.

 The above analysis confirms (\ref{cafsusy}).  It should also pointed out that massive IIA supergravity does not have a maximally supersymmetric solution
  while all the maximally supersymmetric solutions of standard IIA supergravity are locally isometric to $\bR^{9,1}$ with vanishing fluxes  and constant dilaton \cite{maxsusy}. This in particular
  implies that  $N$ is further restricted. The results have been summarized in table 2.

\begin{table}
\centering
\fontencoding{OML}\fontfamily{cmm}\fontseries{m}\fontshape{it}\selectfont
\begin{tabular}{|c|c|}\hline
${\mathbb{R}}^{n-1,1}\times_w M^{10-n}$& $N$
 \\
\hline\hline

$n=2$&$ N< 31$
\\
\hline
$n=3$&$2k, k\leq 15$
\\
\hline
$n=4$&$4k, k\leq 15$
\\
\hline
$n=5$&$8, 16, 24$
\\
\hline
$n=6$&$8, 16, 24$
\\
\hline
$n= 7, 8,9$&$16$
\\
\hline
$n= 10$&$32$
\\
\hline
\end{tabular}
\label{tab2}
\begin{caption}
{\small {\rm ~~The number of supersymmetries $N$ of ${\mathbb{R}}^{1,1}\times_w M^{10-n}$ is not a priori an even number. The corresponding statement for $AdS_2$ backgrounds
is proven using global considerations which are not applicable in this case.  For the rest, the counting of supersymmetries follows from the properties of KSEs
 and the classification results of \cite{maxsusy, bandos}.  Furthermore,  if the Killing spinors do not depend on ${\mathbb{R}}^{n-1,1}$ coordinates, then all backgrounds with $N>16$
 are locally isometric to $\bR^{9,1}$ with zero fluxes and constant dilaton as a consequence of the homogeneity conjecture \cite{josehom}.
}}
\end{caption}
\end{table}

\section{On the factorization of Killing spinors}

\subsection{AdS backgrounds}

Having solved the KSEs of $\mathrm{AdS}_n\times_w M^{10-n}$ backgrounds without any assumptions on the form of the Killing spinors, one can address the question of
whether  the Killing spinors of these spaces factorize as $\epsilon=\psi\otimes \xi$ where $\psi$ is a Killing spinor on $\mathrm{AdS}_n$ and $\xi$
is a Killing spinor on $ M^{10-n}$. In particular, $\psi$ is assumed to satisfy a KSE of the type
\bea
\nabla_\mu \psi+ \lambda \gamma_\mu \psi=0~,
\eea
where $\nabla$ is the spin connection of $\mathrm{AdS}_n$ and $\lambda$ is a constant related to the radius of  $\mathrm{AdS}_n$.
This is an assumption which has been extensively used in the literature.

This issue has already been addressed in \cite{mads} and \cite{iibads} for the $\mathrm{AdS}_n$  backgrounds of D=11 and IIB supergravities.
In particular, it has been found that such a factorization does not occur.
In addition if one insists on such a factorization, then one gets the incorrect counting for the supersymmetries of well-known backgrounds like
$\mathrm{AdS}_5\times S^5$ and $\mathrm{AdS}_7\times S^4$. The same applies for the backgrounds of (massive) IIA supergravity we have investigated here. After an analysis similar
to the one which has been performed in \cite{mads} and \cite{iibads}, one finds that the  Killing spinors we have found do not factorize
into Killing spinors on $\mathrm{AdS}_n$ and Killing spinors on  $ M^{10-n}$.

\subsection{Flat backgrounds}

The issue of factorization of Killing spinors for $\bR^{n-1,1}\times_w M^{10-n}$ backgrounds is closely related to whether the Killing spinors $\epsilon$ we have found exhibit
a linear dependence on the $\bR^{n-1,1}$ coordinates.  This is because if the Killing spinors factorize, then they should not depend on the coordinates
of $\bR^{n-1,1}$ for the chosen coordinate system.  As $\sigma_\pm$ must lie in the Kernel of $(\Xi_\pm)^2$ as a consequence of integrability conditions, the Killing spinors $\epsilon$ exhibit a $\bR^{n-1,1}$ coordinate dependence, iff
$\sigma_\pm\notin \mathrm{Ker}\, \Xi_\pm$. In many examples we have investigated,  $\sigma_\pm\in\mathrm{Ker}\, (\Xi_\pm)^2$ implies that $\sigma_\pm\in \mathrm{Ker}\, \Xi_\pm$ and so
the Killing spinors $\epsilon$ do not depend on the coordinates of  $\bR^{n-1,1}$.  However, we have not been able to prove this in general.

Suppose that all Killing spinors do not depend on the coordinates of $\bR^{n-1,1}$. If $N>16$, the homogeneity conjecture \cite{josehom} applied on the KSEs on $M^{10-n}$ implies that
$M^{10-n}$ is homogenous space and all the fields are invariant. In particular, $A$ and $\Phi$ are constant. Then the field equations of $A$ and $\Phi$ imply that for all
such backgrounds the fluxes vanish. As a consequence all such backgrounds with $N>16$ are locally isometric to $\bR^{9,1}$ with zero fluxes and constant dilaton.

\section{Conclusions}

We have solved the KSEs of all warped flux  $\mathrm{AdS}_n\times_w M^{10-n}$ and $\bR^{n-1,1}\times_w M^{10-n}$ backgrounds without making any assumptions
on the form of the fields and on that of the Killing spinors apart from imposing the symmetries of $\mathrm{AdS}_n$ and $\bR^{n-1,1}$ on the former, respectively.
This has allowed us to a priori  count the number of supersymmetries preserved by these backgrounds for all $n$, and to identify the independent KSEs that
have to be satisfied on $M^{10-n}$.

Furthermore, we have demonstrated that the Killing spinors of $\mathrm{AdS}_n\times_w M^{10-n}$ can be identified with the zero modes of a Dirac-like
operator on $M^{10-n}$. For this we have demonstrated a new class of Lichnerowicz type theorems utilizing the maximum principle.

We have also explored several other properties like the factorization of Killing spinors of $\mathrm{AdS}_n\times_w M^{10-n}$ and $\bR^{n-1,1}\times_w M^{10-n}$ backgrounds
into Killing spinors on $\mathrm{AdS}_n$ and $\bR^{n-1,1}$, respectively, and Killing spinors on $M^{10-n}$. We have found that in the former case
the Killing spinors do not factorize in such a way.

The identification of fractions of supersymmetry preserved by the $\mathrm{AdS}_n\times_w M^{10-n}$ and $\bR^{n-1,1}\times_w M^{10-n}$ backgrounds is a step
forward towards their classification. It is not a priori obvious that there will be solutions for each allowed fraction of supersymmetry. It is known that
there are several no-go theorems. For example, the (massive) IIA supergravity does not admit maximally supersymmetric $\mathrm{AdS}_n\times_w M^{10-n}$
backgrounds \cite{maxsusy}. It is expected that similar theorems will hold for other fractions.  A related problem is to identify the geometry of the
 $M^{10-n}$ spaces in each case.

 The results of this paper together with those presented in \cite{mads} and \cite{iibads} provide a complete
 picture of the fractions of supersymmetry preserved by flux warped  $\mathrm{AdS}_n$ and $\bR^{n-1,1}$ backgrounds in both ten and eleven dimensions,
 as well as some of their global properties which include new  Lichnerowicz type theorems. It is expected that the systematic exploration of these backgrounds
 for each allowed fraction of supersymmetry   will have
 applications in flux compactification, AdS/CFT, string and M-theory.

\vskip 0.8 cm

\noindent{\bf Acknowledgements} \vskip 0.1cm
GP is partially supported by the STFC grant ST/J002798/1.
JG is supported by the STFC grant, ST/1004874/1.
JG would like to thank the
Department of Mathematical Sciences, University of Liverpool for hospitality during which part of this work
was completed.

\vskip 1cm

\appendix

\section{Conventions}

\subsection{Form and spinor conventions}

Our form conventions are as follows. Let $\omega$ be a k-form, then
\bea
\omega={1\over k!} \omega_{i_1\dots i_k} dx^{i_1}\wedge\dots \wedge dx^{i_k}~,
\eea
and
\bea
d\omega={1\over k!} \partial_{i_1} \omega_{i_2\dots i_{k+1}} dx^{i_1}\wedge\dots \wedge dx^{i_{k+1}}~,
\eea
leading to
\bea
(d\omega)_{i_1\dots i_{k+1}}= (k+1) \partial_{[i_1} \omega_{i_2\dots i_{k+1}]}~.
\eea

Furthermore, we write
\bea
\omega^2= \omega_{i_1\dots i_k} \omega^{i_1\dots i_k}~,~~~\omega^2_{i_1 i_{2}}=\omega_{i_1j_1\dots j_{k-1}} \omega_{i_2}{}^{j_1\dots j_{k-1}} \ .
\eea
Given a volume form $d\mathrm{vol}={1\over n!} \epsilon_{i_1\dots i_n} dx^{i_1}\wedge \dots \wedge dx^{i_n}$, the Hodge dual of $\omega$ is defined as
\bea
*\omega\wedge\chi = (\chi, \omega)\, d\mathrm{vol}
\eea
where
\bea
(\chi, \omega)={1\over k!} \chi_{i_1\dots i_k} \omega^{i_1\dots i_k}~.
\eea
So
\bea
*\omega_{i_1\dots i_{n-k}}={1\over k!} \epsilon_{i_1\dots i_{n-k}}{}^{j_1\dots j_k} \omega_{j_1\dots j_k}~.
\eea

It is well-known that for every form $\omega$, one can define a Clifford algebra element ${\slashed \omega}$ given by
\bea
{\slashed\omega}=\omega_{i_1\dots i_k} \Gamma^{i_1\dots i_k}~,
\eea
where $\Gamma^i$, $i=1,\dots n$, are the Dirac gamma matrices. In addition we introduce the notation
\bea
{\slashed\omega}_{i_1}= \omega_{i_1 i_2 \dots i_k} \Gamma^{i_2\dots i_k}~,~~~\Gamma\slashed{\omega}_{i_1}= \Gamma_{i_1}{}^{
i_2\dots i_{k+1}} \omega_{i_2\dots i_{k+1}}~,
\eea
as it is helpful in many of the expressions we have presented.

\subsection{IIA supergravity conventions}

Our conventions are close to those of \cite{howe} and {\cite{iiaclass}. The bosonic fields of  IIA supergravity are the metric $g$, a 2-form field strength $F$, a 3-form field strength $H$,
  a 4-form field strength $G$, the dilaton $\Phi$ and the massive IIA supergravity has another scalar $S$. In particular, the KSEs of (massive) IIA supergravity are the vanishing
conditions of
\bea
{\cal D}_M \epsilon &\equiv& \nabla_M\epsilon + \tfrac{1}{8} \slashed{H}_{M}\Gamma_{11}\epsilon +\tfrac{1}{8}  S\Gamma_M \epsilon
 +\tfrac{1}{16} \slashed {F}\Gamma_M \Gamma_{11} \epsilon +\tfrac{1}{8\cdot 4!} \slashed {G}\Gamma_M \epsilon ~,
\cr
{\cal A} \epsilon &\equiv & \slashed{d}\Phi \epsilon + \tfrac{1}{12} \slashed{H}\Gamma_{11}\epsilon +\tfrac{5}{4}  S \epsilon  +\tfrac{3}{8}\slashed{F}\Gamma_{11}\epsilon +\tfrac{1}{4\cdot 4!} \slashed{G} \epsilon ~.
\eea
Furthermore, the  field equations and Bianchi identities of (massive) IIA supergravity  are
\bea
&&  R_{MN} -\frac{1}{12}  G^2_{MN}  +\frac{1}{96} g_{MN} G^2 + \frac{1}{4}g_{MN}  S^2 -\frac{1}{4}H^2_{MN} -\frac{1}{2}  F^2_{MN}
\cr
&&~~~~~~ + \frac{1}{8} g_{MN}  F^2 +2 \nabla_M \partial_N \Phi=0~,
\cr
&& \nabla^2\Phi -2 (d\Phi)^2-\frac{3}{8} F^2 -\frac{1}{96} G^2  +\frac{1}{12} H^2 -\frac{5}{4} S{}^2=0~,
\cr
&&   \nabla^P H_{MNP} - 2 (\partial^P \Phi) H_{MNP} -\frac{1}{2}  G_{MNP_1P_2}  F^{P_1P_2}
\cr
&& \qquad\qquad  -  F_{MN}  S+\frac{1}{48} *G_{MNP_1\dots P_4}  G^{P_1\dots P_4}=0 ~,
\cr
&&  \nabla^P  F_{MP} - (\partial^P \Phi) F_{MP} + \frac{1}{6}  G_{MP_1P_2P_3}H^{P_1P_2P_3}=0~,
\cr
&&  \nabla^P  G_{M_1 M_2 M_3 P} - (\partial^P \Phi)  G_{M_1 M_2 M_3 P}
\cr
&&~~~~~~~-\frac{1}{6}*G_{M_1 M_2 M_3 P_1P_2P_3}H^{P_1P_2P_3}  =0~,
\eea
and
\bea
&& dS= Sd\Phi~,~~~ dH=0 ~, ~~~dF- d\Phi\wedge  F-H  S=0 ~,
\cr
&& d G -d\Phi\wedge  G-  F\wedge H=0~.
\eea
Note that the first Bianchi identity can be solved as $S= e^\Phi m$, where $m$ is a constant which is related to the cosmological constant
of massive IIA supergravity.


\section{$AdS_3$ Solutions}
\label{ads3co}

\subsection{$AdS_n$ backgrounds}

Before proceeding with the $AdS_3$ analysis, it is useful to consider
some aspects which are common to all $AdS_n$ solutions.
To prove the formula (\ref{maxprii}) for $AdS_n$ backgrounds, one defines
\begin{equation}
 \hat{\nabla}^{(+, q_1, q_2)}_{i} = \nabla^{( + )}_{i} + q_1 A^{-1} \Gamma_{z i} \mathbb{B}^{( + )} + q_2 \Gamma_i \mathcal{A}^{(+)} ,
\end{equation}
where $q_1, q_2$ are constants which will be specified later.  Next introduce the associated Dirac operator
\bea
{\mathscr D}^{(+,q_1, q_2)}=  \Gamma^i \hat{\nabla}^{(+, q_1, q_2)}_i ~,
\eea
and write for convenience
\begin{equation}
 \mathbb{A}^{( +, q_1, q_2 )} = -q_1 A^{-1} \Gamma_z \mathbb{B}^{( + )} + q_2 \mathcal{A}^{(+)}~.
\end{equation}
Furthermore, we assume that
\begin{equation}
 {\mathscr D}^{(+,q_1, q_2)} \chi_+ = \left( \Gamma^i \nabla_i + \Gamma^i \Psi^{(+)}_i + (10-n) \mathbb{A}^{(+, q_1, q_2)} \right) \chi_+ = 0~.
 \label{conads3x}
\end{equation}
Next one expands the Laplacian as
\begin{equation}
 \nabla^2 \left\| \chi_+ \right\| ^2 = 2 \left\| \nabla \chi_+ \right\| ^2 + 2 \left\langle \chi_+, \nabla^2 \chi_+ \right\rangle~.
\end{equation}
The first term can be evaluated to find
\begin{align} \nonumber
 2 \left\| \nabla \chi_+ \right\| ^2 &= 2 \left\| \hat{\nabla}^{(+, q_1, q_2)} \chi_+ \right\| ^2 - 4 \left\langle \chi_+, \left( \Psi^{( + ) i \dagger} + \mathbb{A}^{(+, q_1, q_2) \dagger} \Gamma^i \right) \nabla_i \chi_+ \right\rangle
 \\ \nonumber
 & \bigindent - 2 \left\langle \chi_+, \left( \Psi^{( + ) i \dagger} + \mathbb{A}^{(+, q_1, q_2) \dagger} \Gamma^i \right) \left( \Psi^{( + )}_i + \Gamma_i \mathbb{A}^{(+, q_1, q_2)} \right) \chi_+ \right\rangle
 \\ \nonumber
 &= 2 \left\| \hat{\nabla}^{(+, q_1, q_2)} \chi_+ \right\| ^2 - 4 \left\langle \chi_+, \Psi^{( + ) i \dagger} \nabla_i \chi_+ \right\rangle
 \\ \nonumber
 & \bigindent - 2 \left\langle \chi_+, \left( \Psi^{( + ) i \dagger} - \mathbb{A}^{(+, q_1, q_2) \dagger} \Gamma^i \right) \left( \Psi^{( + )}_i + \Gamma_i \mathbb{A}^{(+, q_1, q_2)} \right) \chi_+ \right\rangle~,
\end{align}
while the second term can be rearranged as
\begin{align}
2 &\left\langle \chi_+, \nabla^2 \chi_+ \right\rangle = 2 \left\langle \chi_+, \Gamma^i \nabla_i \left( \Gamma^j \nabla_j \chi_+ \right) \right\rangle + \frac{1}{2} R^{(10-n)} \left\| \chi_+ \right\| ^2
 \\ \nonumber
 &= \frac{1}{2} R^{(10-n)} \left\| \chi \right\| ^2 - 2 \left\langle \chi_+, \nabla_i \left( \Gamma^i \Gamma^j \Psi^{( + )}_j + (10-n) \Gamma^i \mathbb{A}^{(+, q_1, q_2)} \right) \chi_+ \right\rangle
 \\ \nonumber
 & \bigindent - 2 \left\langle \chi_+, \left( \Gamma^i \Gamma^j \Psi^{( + )}_j + (10-n) \Gamma^i \mathbb{A}^{(+, q_1, q_2)} \right) \nabla_i \chi_+ \right\rangle~,
\end{align}
upon  using $\nabla^2=\slashed{\nabla}^2+{1\over4} R^{(10-n)}$ and (\ref{conads3x}).
Thus, we obtain
\begin{align}  \label{eq:massless_laplacian_expansion}
& \nabla^2 \left\| \chi_+ \right\| ^2 = 2 \left\| \hat{\nabla}^{( +, q_1, q_2 )} \chi_+ \right\| ^2 + \frac{1}{2} R^{(10-n)} \left\| \chi_+ \right\| ^2
 \\ \nonumber
 & ~~~ + \left\langle \chi_+, \left[ -4 \Psi^{(+) i \dagger} - 2 \Gamma^i \Gamma^j \Psi^{(+)}_j - 2 (10-n) q_1 A^{-1} \Gamma^{z i} \mathbb{B}^{(+)} \right. \right.
 \\ \nonumber
 &~~~ \left. \left. - 2 (10-n) q_2 \Gamma^i \mathcal{A}^{(+)} \right] \nabla_i \chi_+ \right\rangle
 \\ \nonumber
 &~~~ + \left\langle \chi_+, -2 \left( \Psi^{(+) i \dagger} - \mathbb{A}^{(+, q_1, q_2)  \dagger} \Gamma^i \right) \left( \Psi^{(+)}_i + \Gamma_i \mathbb{A}^{(+, q_1, q_2)} \right) \chi_+ \right\rangle
 \\ \nonumber
 &~~~ + \left\langle \chi_+, \nabla_i \left[ -2 \Gamma^i \Gamma^j \Psi^{(+)}_j - 2 (10-n) q_1 A^{-1} \Gamma^{z i} \mathbb{B}^{(+)}  - 2 (10-n) q_2 \Gamma^i \mathcal{A}^{(+)} \right] \chi_+ \right\rangle~.
\end{align}
Note that $\Gamma_i^\dagger =\Gamma_i$ and $\Gamma_{11}^\dagger=\Gamma_{11}$ and so $\Gamma_{ij}^\dagger=-\Gamma_{ij}$.
From here on, the computation depends on $n$ and it will be explained in each case separately.

\subsection{$AdS_3$: Standard IIA solutions with $S=0$}

First, we  write the third term of (\ref{eq:massless_laplacian_expansion}) in the form $ \alpha^i \nabla_i \left\| \chi_+ \right\| ^2 + \left\langle \chi_+, \mathcal{F} \slashed{\nabla} \chi_+ \right\rangle $ for some Clifford algebra element $\mathcal{F}$ which depends on the fields. Expressing  this term in terms  of the fields, and setting $n=3$ throughout, one finds
\begin{align}
 & \reverseindent \left\langle \chi_+, \left[ -4 \Psi^{(+) i \dagger} - 2 \Gamma^i \Gamma^j \Psi^{(+)}_{j} - 14 q_1 A^{-1} \Gamma^{z i} \mathbb{B}^{(+)} - 14 q_2 \Gamma^i \mathcal{A}^{(+)} \right] \nabla_{i} \chi_+ \right\rangle
 \\ \nonumber
 &= \left\langle \chi_+, \left[ \frac{7 q_1 c}{\ell} A^{-1} \Gamma^{z i} - \left[ 3 + 7 q_1 \right] A^{-1} \partial^i A - \left[ 1 + 7 q_1 \right] A^{-1} \left( \Gamma \slashed{\partial} A \right)^i  \right. \right.
 \\ \nonumber
 & \bigindent \bigindent - 14 q_2 \Gamma^i \slashed{\partial} \Phi - \frac{1 + 14 q_2}{4} \slashed{Z}^i \Gamma_{11} - \frac{3 + 14 q_2}{12} \Gamma \slashed{Z}^i \Gamma_{11}
 \\ \nonumber
 & \bigindent \bigindent - \frac{7 q_1 + 14 q_2}{2} W \Gamma^{z i} \Gamma_{11} - \frac{5 + 7 q_1 + 42 q_2}{8} \Gamma^i \slashed{F} \Gamma_{11}
 \\ \nonumber
 & \bigindent \bigindent \left. \left. + \frac{-1 - 7 q_1 - 14 q_2}{96} \Gamma^i \slashed{Y}
 +{-3 +7q_1-14 q_2 \over 4}\Gamma^i \slashed{X} \Gamma_z
  \right] \nabla_i \chi_+ \right\rangle~.
\end{align}
Then it can be rewritten in the form required provided that one chooses  $ q_2 = -\frac{1}{7} $ and $ q_1 = \frac{1}{7} $. In particular,
\begin{align}
 & \reverseindent \left\langle \chi_+, \left[ -4 \Psi^{(+) i \dagger} - 2 \Gamma^i \Gamma^j \Psi^{(+)}_{j} - 2 A^{-1} \Gamma^{z i} \mathbb{B}^{(+)} + 2 \Gamma^i \mathcal{A}^{(+)} \right] \nabla_{i} \chi_+ \right\rangle
 \\ \nonumber
 &= \left\langle \chi_+, \left[ \frac{c}{\ell} A^{-1} \Gamma_{z i} - 4 A^{-1} \partial_i A - 2 A^{-1} \left( \Gamma \slashed{\partial} A \right) _i + 2 \Gamma_i \slashed{\partial} \Phi \right. \right.
 \\ \nonumber
 & \bigindent \bigindent \left. \left. - \frac{1}{12} \left( \Gamma \slashed{Z} \right) _i \Gamma_{11} + \frac{1}{4} \slashed{Z}_i \Gamma_{11} + \frac{1}{2} W \Gamma_{z i} \Gamma_{11} \right] \nabla^i \chi_+ \right\rangle~,
\end{align}
for
\begin{equation}
 \mathcal{F} = \frac{c}{\ell} A^{-1} \Gamma_z + 2 A^{-1} \slashed{\partial} A - 2 \slashed{\partial} \Phi - \frac{1}{12} \slashed{Z} \Gamma_{11} - \frac{1}{2} W \Gamma_z \Gamma_{11}~,
\end{equation}
and $ \alpha_i = -3 A^{-1} \partial_i A + 2 \partial_i \Phi $.

Combining the ${\cal{F}}$ term with the fourth term in \eqref{eq:massless_laplacian_expansion}, we find
\begin{align} \nonumber
 & \reverseindent \left\langle \chi_+, -2 \left( \Psi^{( + ) i \dagger} + \frac{1}{7} A^{-1} \mathbb{B}^{(+) \dagger} \Gamma^{z i} + \frac{1}{7} \mathcal{A}^{(+) \dagger} \Gamma^i + \frac{1}{2} \mathcal{F} \Gamma^i \right) \left( \Psi^{( + )}_i + \frac{1}{7} A^{-1} \Gamma_{z i} \mathbb{B}^{(+)} - \frac{1}{7} \Gamma_i \mathcal{A}^{(+)} \right) \chi_+ \right\rangle
 \\ \nonumber
 &= \left\langle \chi_+, -2 \left[ \frac{3 c}{7 \ell} A^{-1} \Gamma^{z i} + \frac{10}{7} A^{-1} \partial^i A - \frac{13}{14} A^{-1} \Gamma \slashed{\partial}^i A - \frac{6}{7} \slashed{\partial} \Phi \Gamma^i - \frac{1}{28} \slashed{Z}^i \Gamma_{11} - \frac{5}{168} \Gamma \slashed{Z}^i \Gamma_{11} \right. \right.
 \\ \nonumber
 & \bigindent \bigindent + \frac{3}{14} W \Gamma^{z i} \Gamma_{11} + \frac{3}{28} \Gamma \slashed{F}^i \Gamma_{11} + \frac{1}{28} \slashed{F}^i \Gamma_{11} + \frac{1}{168} \Gamma \slashed{Y}^i + \frac{1}{56} \slashed{Y}^i
 \\ \nonumber
 & \bigindent \bigindent \left. - \frac{5}{28} \Gamma \slashed{X}^i \Gamma_z - \frac{1}{14} X^i \Gamma_z \right]
 \\ \nonumber
 & \bigindent \times \left[ -\frac{c}{14 \ell} A^{-1} \Gamma_{z i} + \frac{4}{7} A^{-1} \partial_i A + \frac{1}{14} A^{-1} \Gamma \slashed{\partial}_i A - \frac{1}{7} \Gamma_i \slashed{\partial} \Phi + \frac{5}{56} \slashed{Z}_i \Gamma_{11} - \frac{1}{84} \Gamma \slashed{Z}_i \Gamma_{11} \right.
 \\ \nonumber
 & \bigindent \bigindent - \frac{1}{28} W \Gamma_{z i} \Gamma_{11} + \frac{1}{56} \Gamma \slashed{F}_i \Gamma_{11} - \frac{3}{14} \slashed{F}_i \Gamma_{11} + \frac{1}{224} \Gamma \slashed{Y}_i - \frac{1}{42} \slashed{Y}_i
 \\ \nonumber
 & \bigindent \bigindent \left. \left. + \frac{1}{14} \Gamma \slashed{X}_i \Gamma_z - \frac{5}{28} X_i \Gamma_z \right] \chi_+ \right\rangle
 \\ \nonumber
 &= \left\langle \chi_+, \left[ -\frac{3}{7 \ell^2} A^{-2} - \frac{17}{7} A^{-2} (dA)^2 + \frac{26}{7} A^{-1} \partial^i A \partial_i \Phi - \frac{12}{7} (d\Phi)^2 + \frac{c}{42 \ell} A^{-1} \slashed{Z} \Gamma_z \Gamma_{11} \right. \right.
 \\ \nonumber
 & \bigindent - \frac{1}{42} A^{-1} \partial_i A \Gamma \slashed{Z}^i \Gamma_{11} + \frac{1}{21} \partial_i \Phi \Gamma \slashed{Z}^i \Gamma_{11} - \frac{1}{504} \slashed{Z} \slashed{Z} - \frac{1}{24} Z^2 - \frac{3 c}{7 \ell} A^{-1} W \Gamma_{11}
 \\ \nonumber
 & \bigindent + \frac{1}{84} W \slashed{Z} \Gamma_z  - \frac{3}{28} W^2 + \frac{c}{28 \ell} A^{-1} \slashed{F} \Gamma_z \Gamma_{11} + \frac{1}{28} A^{-1} \partial_i A \Gamma \slashed{F}^i \Gamma_{11}
 \\ \nonumber
 & \bigindent - \frac{1}{14} \partial_i \Phi \Gamma \slashed{F}^i \Gamma_{11} - \frac{3}{112} \slashed{Z} \slashed{F} + \frac{1}{8} \slashed{Z}_i \slashed{F}^i - \frac{1}{28} \slashed{F} \slashed{F} - \frac{1}{8} F^2  + \frac{c}{112 \ell} A^{-1} \slashed{Y} \Gamma_z
 \\ \nonumber
 & \bigindent + \frac{1}{112} A^{-1} \partial_i A \Gamma \slashed{Y}^i - \frac{1}{56} \partial_i \Phi \Gamma \slashed{Y}^i - \frac{13}{42 \cdot 96} \slashed{Z} \slashed{Y} \Gamma_{11} + \frac{1}{48} \slashed{Z}_i \slashed{Y}^i \Gamma_{11} - \frac{1}{448} \slashed{F} \slashed{Y} \Gamma_{11}
 \\ \nonumber
 & \bigindent + \frac{1}{168 \cdot 96} \slashed{Y} \slashed{Y} - \frac{1}{96} Y^2 + \frac{5 c}{14 \ell} A^{-1} \slashed{X} - \frac{5}{14} A^{-1} X_i \partial^i A \Gamma_z + \frac{5}{7} X_i \partial^i \Phi \Gamma_z
 \\ \label{eq:massless_algebraic_product}
 & \bigindent \left. \left. + \frac{3}{56} X_i \slashed{Z}^i \Gamma_z \Gamma_{11} - \frac{1}{28} X_i \Gamma \slashed{F}^i \Gamma_z \Gamma_{11} + \frac{1}{168} X_i \slashed{Y}^i \Gamma_z - \frac{5}{28} X^2 \right] \chi_+ \right\rangle~.
\end{align}

Using the field equations and Bianchi identities, one can express the last term of \eqref{eq:massless_laplacian_expansion} as
\begin{align} \nonumber
 & \reverseindent \left\langle \chi_+, \nabla_i \left[ -2 \Gamma^i \Gamma^j \Psi^{( + )}_j - 2 A^{-1} \Gamma^{z i} \mathbb{B}^{(+)} + 2 \Gamma^i \mathcal{A}^{(+)} \right] \chi_+ \right\rangle
 \\ \nonumber
 &= \left\langle \chi_+, \left[ -2 \nabla^2 \ln A + 2 \nabla^2 \Phi - \frac{1}{48} \slashed{dZ} \Gamma_{11} + \frac{1}{12} \slashed{dF} \Gamma_{11} + \frac{1}{240} \slashed{dY} - \frac{1}{2} \nabla_i X^i \Gamma_z \right] \chi_+ \right\rangle
 \\ \nonumber
 &= \left\langle \chi_+, \left[ \frac{4}{\ell^2} A^{-2} + 6 A^{-2} (dA)^2 - 10 A^{-1} \partial^i A \partial_i \Phi + 4 (d\Phi)^2 - \frac{1}{6} Z^2 + \frac{1}{4} \partial_i \Phi \Gamma \slashed{F}^i \Gamma_{11} \right. \right.
 \\ \nonumber
 & \bigindent ~~~ + \frac{1}{24} \slashed{Z} \slashed{F} - \frac{1}{8} \slashed{Z}^i \slashed{F}_i + \frac{1}{2} F^2 + \frac{1}{48} \partial_i \Phi \Gamma \slashed{Y}^i
 \\ \label{eq:massless_field_eqns_and_bianchis}
 & \bigindent ~~~ \left. \left. + \frac{1}{288} \slashed{Z} \slashed{Y} \Gamma_{11} - \frac{1}{48} \slashed{Z}_i \slashed{Y}^i \Gamma_{11} - \frac{1}{2} X^i \partial_i \Phi \Gamma_z - X^2 \right] \chi_+ \right\rangle
\end{align}
while using the field equations again the term in \eqref{eq:massless_laplacian_expansion} containing the Ricci scalar  can be written as
\begin{align} \nonumber
 & \reverseindent \frac{1}{2} R^{(7)} \left\| \chi_+ \right\| ^2 = \left\langle \chi_+, \left[ -\frac{3}{\ell^2} A^{-2} - 3 A^{-2} (dA)^2 + 6 A^{-1} \partial^i A \partial_i \Phi - 2 (d\Phi)^2 \right. \right.
 \\ \label{eq:massless_curvature}
 & \left. \left. + \frac{5}{24} Z^2 + \frac{1}{4} W^2 - \frac{3}{8} F^2 + \frac{1}{96} Y^2 + \frac{5}{4} X^2 \right] \chi_+ \right\rangle .
\end{align}
Using \eqref{eq:massless_algebraic_product}, \eqref{eq:massless_field_eqns_and_bianchis}, and \eqref{eq:massless_curvature}, the right-hand-side of \eqref{eq:massless_laplacian_expansion}, apart from the first term and the $\alpha^i \partial_i \parallel \chi_+ \parallel^2$ term,
can be evaluated as
\begin{align} \nonumber
 & \reverseindent \left\langle \chi_+, \left[ \frac{4}{7 \ell^2} A^{-2} + \frac{4}{7} A^{-2} (dA)^2 - \frac{2}{7} A^{-1} \partial^i A \partial_i \Phi + \frac{2}{7} (d\Phi)^2 + \frac{c}{42 \ell} A^{-1} \slashed{Z} \Gamma_z \Gamma_{11} \right. \right.
 \\ \nonumber
 & - \frac{1}{42} A^{-1} \partial_i A \Gamma \slashed{Z}^i \Gamma_{11} + \frac{1}{21} \partial_i \Phi \Gamma \slashed{Z}^i \Gamma_{11} - \frac{1}{504} \slashed{Z} \slashed{Z} - \frac{3 c}{7 \ell} A^{-1} W \Gamma_{11}
 \\ \nonumber
 & + \frac{1}{84} W \slashed{Z} \Gamma_z  + \frac{1}{7} W^2 + \frac{c}{28 \ell} A^{-1} \slashed{F} \Gamma_z \Gamma_{11} + \frac{1}{28} A^{-1} \partial_i A \Gamma \slashed{F}^i \Gamma_{11}
 \\ \nonumber
 & + \frac{5}{28} \partial_i \Phi \Gamma \slashed{F}^i \Gamma_{11} + \frac{5}{336} \slashed{Z} \slashed{F} - \frac{1}{28} \slashed{F} \slashed{F} + \frac{c}{112 \ell} A^{-1} \slashed{Y} \Gamma_z + \frac{1}{112} A^{-1} \partial_i A \Gamma \slashed{Y}^i
 \\ \nonumber
 & + \frac{1}{336} \partial_i \Phi \Gamma \slashed{Y}^i + \frac{1}{42 \cdot 96} \slashed{Z} \slashed{Y} \Gamma_{11} - \frac{1}{448} \slashed{F} \slashed{Y} \Gamma_{11} + \frac{1}{168 \cdot 96} \slashed{Y} \slashed{Y}
 \\ \nonumber
 & + \frac{5 c}{14 \ell} A^{-1} \slashed{X} - \frac{5}{14} A^{-1} X_i \partial^i A \Gamma_z + \frac{3}{14} X_i \partial^i \Phi \Gamma_z
 \\
 & \left. \left. + \frac{3}{56} X_i \slashed{Z}^i \Gamma_z \Gamma_{11} - \frac{1}{28} X_i \Gamma \slashed{F}^i \Gamma_z \Gamma_{11} + \frac{1}{168} X_i \slashed{Y}^i \Gamma_z + \frac{1}{14} X^2 \right] \chi_+ \right\rangle~.
\end{align}

Next we compare the above expression  to
\begin{align} \nonumber
 \left\| \mathbb{B}^{(+)} \chi_+ \right\| ^2 &= \left\langle \chi_+, \left[ \frac{1}{4 \ell^2} + \frac{1}{4} (dA)^2 - \frac{c}{4 \ell} A W \Gamma_{11} + \frac{1}{16} A^2 W^2 + \frac{c}{16 \ell} A \slashed{F} \Gamma_z \Gamma_{11} \right. \right.
 \\ \nonumber
 & ~~~ + \frac{1}{16} A \partial_i A \Gamma \slashed{F}^i \Gamma_{11} - \frac{1}{256} A^2 \slashed{F} \slashed{F} + \frac{c}{192 \ell} A \slashed{Y} \Gamma_z + \frac{1}{192} A \partial_i A  \Gamma \slashed{Y}^i
 \\ \nonumber
 & ~~~- \frac{1}{16 \cdot 96} A^2 \slashed{F} \slashed{Y} \Gamma_{11} + \frac{1}{192^2} A^2 \slashed{Y} \slashed{Y} + \frac{c}{8 \ell} A \slashed{X} - \frac{1}{8} A X^i \partial_i A \Gamma_z
 \\
 &~~~ \left. \left. + \frac{1}{64} A^2 X_i \Gamma \slashed{F}^i \Gamma_z \Gamma_{11} + \frac{1}{192} A^2 X_i \slashed{Y}^i \Gamma_z + \frac{1}{64} A^2 X^2 \right] \chi_+ \right\rangle~,
\end{align}

\begin{align} \nonumber
 \left\langle \Gamma_z \mathbb{B}^{(+)} \chi_+, \mathcal{A}^{(+)} \chi_+ \right\rangle &= \left\langle \chi_+, \left[ -\frac{1}{2} \partial^i A \partial_i \Phi + \frac{c}{24 \ell} \slashed{Z} \Gamma_z \Gamma_{11} - \frac{1}{24} \partial_i A \Gamma \slashed{Z}^i \Gamma_{11} + \frac{c}{4 \ell} W \Gamma_{11} \right. \right.
 \\ \nonumber
 & \bigindent - \frac{1}{48} A W \slashed{Z} \Gamma_z - \frac{1}{8} A W^2 - \frac{3c}{16 \ell} \slashed{F} \Gamma_z \Gamma_{11} - \frac{3}{16} \partial_i A \Gamma \slashed{F}^i \Gamma_{11}
 \\ \nonumber
 & \bigindent - \frac{1}{16} A \partial_i \Phi \Gamma \slashed{F}^i \Gamma_{11} - \frac{A}{192} \slashed{Z} \slashed{F} + \frac{3}{128} A \slashed{F} \slashed{F} - \frac{c}{192 \ell} \slashed{Y} \Gamma_z
 \\ \nonumber
 & \bigindent - \frac{1}{192} \partial_i A \Gamma \slashed{Y}^i - \frac{1}{192} A \partial_i \Phi \Gamma \slashed{Y}^i - \frac{1}{24 \cdot 96} A \slashed{Z} \slashed{Y} \Gamma_{11}
 \\ \nonumber
 & \bigindent + \frac{1}{384} A \slashed{F} \slashed{Y} \Gamma_{11} - \frac{1}{96 \cdot 192} A \slashed{Y} \slashed{Y} + \frac{c}{8 \ell} \slashed{X} - \frac{1}{8} X^i \partial_i A \Gamma_z
 \\ \nonumber
 & \bigindent + \frac{1}{8} A X^i \partial_i \Phi \Gamma_z + \frac{1}{32} A X_i \slashed{Z}^i \Gamma_z \Gamma_{11}
 \\
 & \bigindent \left. \left. - \frac{1}{32} A X_i \Gamma \slashed{F}^i \Gamma_z \Gamma_{11} + \frac{1}{32} A X^2 \right] \chi_+ \right\rangle~,
\end{align}
and
\begin{align} \nonumber
 \left\| \mathcal{A}^{(+)} \chi_+ \right\| ^2 &= \left\langle \chi_+, \left[ (d\Phi)^2 + \frac{1}{6} \partial_i \Phi \Gamma \slashed{Z}^i \Gamma_{11} - \frac{1}{144} \slashed{Z} \slashed{Z} + \frac{1}{12} W \slashed{Z} \Gamma_z + \frac{1}{4} W^2 \right. \right.
 \\ \nonumber
 &~~ + \frac{3}{4} \partial_i \Phi \Gamma \slashed{F}^i \Gamma_{11} + \frac{1}{16} \slashed{Z} \slashed{F} - \frac{9}{64} \slashed{F} \slashed{F} + \frac{1}{48} \partial_i \Phi \Gamma \slashed{Y}^i
 \\ \nonumber
 &~~+ \frac{1}{576} \slashed{Z} \slashed{Y} \Gamma_{11} - \frac{1}{128} \slashed{F} \slashed{Y} \Gamma_{11} + \frac{1}{96^2} \slashed{Y} \slashed{Y} + \frac{1}{2} X^i \partial_i \Phi \Gamma_z
 \\
 &~~\left. \left. + \frac{1}{8} X_i \slashed{Z}^i \Gamma_z \Gamma_{11} - \frac{3}{16} X_i \Gamma \slashed{F}^i \Gamma_z \Gamma_{11} - \frac{1}{48} X_i \slashed{Y}^i \Gamma_z + \frac{1}{16} X^2 \right] \chi_+ \right\rangle~,
\end{align}
to find that
\begin{align}
 &  \nabla^2 \left\| \chi_+ \right\| ^2 + \left( 3 A^{-1} \partial_i A - 2 \partial_i \Phi \right) \nabla^i \left\| \chi_+ \right\| ^2
 \\ \nonumber
 &= \left\| \hat{\nabla}^{(+)} \chi_+ \right\| ^2 + \frac{16}{7} A^{-2} \left\| \mathbb{B}^{(+)} \chi_+ \right\| ^2 + \frac{4}{7} A^{-1} \left\langle \Gamma_z \mathbb{B}^{(+)} \chi_+, \mathcal{A}^{(+)} \chi_+ \right\rangle + \frac{2}{7} \left\| \mathcal{A}^{(+)} \chi_+ \right\| ^2~,
\end{align}
where $\hat{\nabla}^{( +, {1\over7}, -{1\over7} )}=\hat{\nabla}^{(+)}$.
This proves (\ref{maxprii}), or equivalently (\ref{maxads3}), for the the standard IIA  $\mathrm{AdS}_3$ backgrounds.

\subsection{$AdS_3$: Massive IIA solutions with $W=0$}

Next let us establish (\ref{maxads3}) for the massive IIA $\mathrm{AdS}_3$ backgrounds.
Continuing the computation in the same way as in the previous case, one finds that
\begin{align}
 &  \left\langle \chi_+, \left[ -4 \Psi^{(+) i \dagger} - 2 \Gamma^i \Gamma^j \Psi^{(+)}_{j} - 14 q_1 A^{-1} \Gamma^{z i} \mathbb{B}^{(+)} - 14 q_2 \Gamma^i \mathcal{A}^{(+)} \right] \nabla_{i} \chi_+ \right\rangle\nonumber
 \\ \nonumber
 &= \left\langle \chi_+, \left[ \frac{7 q_1 c}{\ell} A^{-1} \Gamma^{z i} - \left[ 3 + 7 q_1 \right] A^{-1} \partial^i A - \left[ 1 + 7 q_1 \right] A^{-1} \left( \Gamma \slashed{\partial} A \right)^i  \right. \right.
 \\ \nonumber
 & \bigindent  - 14 q_2 \Gamma^i \slashed{\partial} \Phi - \frac{1 + 14 q_2}{4} \slashed{Z}^i \Gamma_{11} - \frac{3 + 14 q_2}{12} \Gamma \slashed{Z}^i \Gamma_{11}
 \\ \nonumber
 & \bigindent - \frac{9 + 7 q_1 + 70 q_2}{4} S \Gamma^i - \frac{5 + 7 q_1 + 42 q_2}{8} \Gamma^i \slashed{F} \Gamma_{11}
 \\ \nonumber
 & \bigindent  \left. \left. + \frac{-1 - 7 q_1 - 14 q_2}{96} \Gamma^i \slashed{Y} +{-3 +7q_1-14 q_2 \over 4}\Gamma^i \slashed{X} \Gamma_z \right] \nabla_i \chi_+ \right\rangle .
\end{align}
which can be written in the form $ \alpha^i \nabla_i \left\| \chi_+ \right\| ^2 + \left\langle \chi_+, \mathcal{F} \Gamma^i \tilde{\nabla}_i \chi_+ \right\rangle $
provided that $ q_2 = -\frac{1}{7} $ and $ q_1 = \frac{1}{7} $. In particular, one has
\begin{align}
 &  \left\langle \chi_+, \left[ -4 \Psi^{(+) i \dagger} - 2 \Gamma^i \Gamma^j \Psi^{(+)}_{j} - 2 A^{-1} \Gamma^{z i} \mathbb{B}^{(+)} + 2 \Gamma^i \mathcal{A}^{(+)} \right] \nabla_{i} \chi_+ \right\rangle \nonumber
 \\ \nonumber
 &= \left\langle \chi_+, \left[ \frac{c}{\ell} A^{-1} \Gamma_{z i} - 4 A^{-1} \partial_i A - 2 A^{-1} \left( \Gamma \slashed{\partial} A \right) _i + 2 \Gamma_i \slashed{\partial} \Phi \right. \right.
 \\ \nonumber
 & \bigindent \bigindent \left. \left. - \frac{1}{12} \left( \Gamma \slashed{Z} \right) _i \Gamma_{11} + \frac{1}{4} \slashed{Z}_i \Gamma_{11} \right] \nabla^i \chi_+ \right\rangle~,
\end{align}
with
\begin{equation}
 \mathcal{F} = \frac{c}{\ell} A^{-1} \Gamma_z + 2 A^{-1} \slashed{\partial} A - 2 \slashed{\partial} \Phi - \frac{1}{12} \slashed{Z} \Gamma_{11}~,
\end{equation}
and $ \alpha_i = -3 A^{-1} \partial_i A + 2 \partial_i \Phi $.

Combining the ${\cal{F}}$ term with the fourth term in (\ref{eq:massless_laplacian_expansion}), we find
\begin{align} \nonumber
 & \reverseindent \left\langle \chi_+, -2 \left( \Psi^{( + ) i \dagger} + \frac{1}{7} A^{-1} \mathbb{B}^{(+) \dagger} \Gamma^{z i} + \frac{1}{7} \mathcal{A}^{(+) \dagger} \Gamma^i + \frac{1}{2} \mathcal{F} \Gamma^i \right) \left( \Psi^{( + )}_i + \frac{1}{7} A^{-1} \Gamma_{z i} \mathbb{B}^{(+)} - \frac{1}{7} \Gamma_i \mathcal{A}^{(+)} \right) \chi_+ \right\rangle
 \\ \nonumber
 &= \left\langle \chi_+, -2 \left[ \frac{3 c}{7 \ell} A^{-1} \Gamma^{z i} + \frac{10}{7} A^{-1} \partial^i A - \frac{13}{14} A^{-1} \Gamma \slashed{\partial}^i A - \frac{6}{7} \slashed{\partial} \Phi \Gamma^i - \frac{1}{28} \slashed{Z}^i \Gamma_{11} - \frac{5}{168} \Gamma \slashed{Z}^i \Gamma_{11} \right. \right.
 \\ \nonumber
 & \bigindent \bigindent + \frac{2}{7} S \Gamma^i + \frac{3}{28} \Gamma \slashed{F}^i \Gamma_{11} + \frac{1}{28} \slashed{F}^i \Gamma_{11} + \frac{1}{168} \Gamma \slashed{Y}^i + \frac{1}{56} \slashed{Y}^i
 \\ \nonumber
 & \bigindent \bigindent \left. - \frac{5}{28} \Gamma \slashed{X}^i \Gamma_z - \frac{1}{14} X^i \Gamma_z \right]
 \\ \nonumber
 & \bigindent \times \left[ -\frac{c}{14 \ell} A^{-1} \Gamma_{z i} + \frac{4}{7} A^{-1} \partial_i A + \frac{1}{14} A^{-1} \Gamma \slashed{\partial}_i A - \frac{1}{7} \Gamma_i \slashed{\partial} \Phi + \frac{5}{56} \slashed{Z}_i \Gamma_{11} - \frac{1}{84} \Gamma \slashed{Z}_i \Gamma_{11} \right.
 \\ \nonumber
 & \bigindent \bigindent - \frac{1}{28} S \Gamma_i + \frac{1}{56} \Gamma \slashed{F}_i \Gamma_{11} - \frac{3}{14} \slashed{F}_i \Gamma_{11} + \frac{1}{224} \Gamma \slashed{Y}_i - \frac{1}{42} \slashed{Y}_i
 \\ \nonumber
 & \bigindent \bigindent \left. \left. + \frac{1}{14} \Gamma \slashed{X}_i \Gamma_z - \frac{5}{28} X_i \Gamma_z \right] \chi_+ \right\rangle
 \\ \nonumber
 &= \left\langle \chi_+, \left[ -\frac{3}{7 \ell^2} A^{-2} - \frac{17}{7} A^{-2} (dA)^2 + \frac{26}{7} A^{-1} \partial^i A \partial_i \Phi - \frac{12}{7} (d\Phi)^2 + \frac{c}{42 \ell} A^{-1} \slashed{Z} \Gamma_z \Gamma_{11} \right. \right.
 \\ \nonumber
 & \bigindent - \frac{1}{42} A^{-1} \partial_i A \Gamma \slashed{Z}^i \Gamma_{11} + \frac{1}{21} \partial_i \Phi \Gamma \slashed{Z}^i \Gamma_{11} - \frac{1}{504} \slashed{Z} \slashed{Z} - \frac{1}{24} Z^2 - \frac{c}{14 \ell} A^{-1} S \Gamma_z
 \\ \nonumber
 & \bigindent - \frac{1}{14} A^{-1} S \slashed{\partial} A + \frac{1}{7} S \slashed{\partial} \Phi - \frac{5}{168} S \slashed{Z} \Gamma_{11} + \frac{1}{7} S^2 + \frac{c}{28 \ell} A^{-1} \slashed{F} \Gamma_z \Gamma_{11}
 \\ \nonumber
 & \bigindent + \frac{1}{28} A^{-1} \partial_i A \Gamma \slashed{F}^i \Gamma_{11} - \frac{1}{14} \partial_i \Phi \Gamma \slashed{F}^i \Gamma_{11} - \frac{3}{112} \slashed{Z} \slashed{F} + \frac{1}{8} \slashed{Z}_i \slashed{F}^i - \frac{1}{28} \slashed{F} \slashed{F} - \frac{1}{8} F^2
 \\ \nonumber
 & \bigindent  + \frac{c}{112 \ell} A^{-1} \slashed{Y} \Gamma_z + \frac{1}{112} A^{-1} \partial_i A \Gamma \slashed{Y}^i - \frac{1}{56} \partial_i \Phi \Gamma \slashed{Y}^i - \frac{13}{42 \cdot 96} \slashed{Z} \slashed{Y} \Gamma_{11}
 \\ \nonumber
 & \bigindent + \frac{1}{48} \slashed{Z}_i \slashed{Y}^i \Gamma_{11} + \frac{1}{168} S \slashed{Y} - \frac{1}{448} \slashed{F} \slashed{Y} \Gamma_{11} + \frac{1}{168 \cdot 96} \slashed{Y} \slashed{Y} - \frac{1}{96} Y^2
 \\ \nonumber
 & \bigindent + \frac{5 c}{14 \ell} A^{-1} \slashed{X} - \frac{5}{14} A^{-1} X_i \partial^i A \Gamma_z + \frac{5}{7} X_i \partial^i \Phi \Gamma_z + \frac{3}{56} X_i \slashed{Z}^i \Gamma_z \Gamma_{11}
 \\ \label{eq:massive_algebraic_product}
 & \bigindent \left. \left. - \frac{1}{28} X_i \Gamma \slashed{F}^i \Gamma_z \Gamma_{11} + \frac{1}{168} X_i \slashed{Y}^i \Gamma_z - \frac{5}{28} X^2 \right] \chi_+ \right\rangle~.
\end{align}

Furthermore using the field equations and Bianchi identities, the last term in (\ref{eq:massless_laplacian_expansion}) can be expressed as
\begin{align} \nonumber
 & \reverseindent \left\langle \chi_+, \nabla_i \left[ -2 \Gamma^i \Gamma^j \Psi^{( + )}_j - 2 A^{-1} \Gamma^{z i} \mathbb{B}^{(+)} + 2 \Gamma^i \mathcal{A}^{(+)} \right] \chi_+ \right\rangle
 \\ \nonumber
 &= \left\langle \chi_+, \left[ -2 \nabla^2 \ln A + 2 \nabla^2 \Phi - \frac{1}{48} \slashed{dZ} \Gamma_{11} + \frac{1}{2} \slashed{\partial} S + \frac{1}{12} \slashed{dF} \Gamma_{11} + \frac{1}{240} \slashed{dY} - \frac{1}{2} \nabla_i X^i \Gamma_z \right] \chi_+ \right\rangle
 \\ \nonumber
 &= \left\langle \chi_+, \left[ \frac{4}{\ell^2} A^{-2} + 6 A^{-2} (dA)^2 - 10 A^{-1} \partial^i A \partial_i \Phi + 4 (d\Phi)^2 - \frac{1}{6} Z^2 + \frac{1}{2} S \slashed{\partial} \Phi \right. \right.
 \\ \nonumber
 & \bigindent \bigindent + \frac{1}{12} S \slashed{Z} \Gamma_{11} + 2 S^2 + \frac{1}{4} \partial_i \Phi \Gamma \slashed{F}^i \Gamma_{11} + \frac{1}{24} \slashed{Z} \slashed{F} - \frac{1}{8} \slashed{Z}^i \slashed{F}_i
 \\ \nonumber
 & \bigindent \bigindent + \frac{1}{2} F^2 + \frac{1}{48} \partial_i \Phi \Gamma \slashed{Y}^i + \frac{1}{288} \slashed{Z} \slashed{Y} \Gamma_{11}
 \\ \label{eq:massive_field_eqns_and_bianchis}
 & \bigindent \bigindent \left. \left. - \frac{1}{48} \slashed{Z}_i \slashed{Y}^i \Gamma_{11} - \frac{1}{2} X^i \partial_i \Phi \Gamma_z - X^2 \right] \chi_+ \right\rangle~,
\end{align}
while the  term involving the Ricci scalar now reads as
\begin{align} \nonumber
 & \reverseindent \frac{1}{2} R^{(7)} \left\| \chi_+ \right\| ^2 = \left\langle \chi_+, \left[ -\frac{3}{\ell^2} A^{-2} - 3 A^{-2} (dA)^2 + 6 A^{-1} \partial^i A \partial_i \Phi - 2 (d\Phi)^2 \right. \right.
 \\ \label{eq:massive_curvature}
 & \left. \left. + \frac{5}{24} Z^2 - \frac{7}{4} S^2 - \frac{3}{8} F^2 + \frac{1}{96} Y^2 + \frac{5}{4} X^2 \right] \chi_+ \right\rangle~.
\end{align}
Using \eqref{eq:massive_algebraic_product}, \eqref{eq:massive_field_eqns_and_bianchis}, and \eqref{eq:massive_curvature}, one finds that the right-hand-side of (\ref{eq:massless_laplacian_expansion}), apart from the first term and the $\alpha^i \partial_i \parallel \chi_+ \parallel^2$ term, can be written as
\begin{align} \nonumber
 & \reverseindent \left\langle \chi_+, \left[ \frac{4}{7 \ell^2} A^{-2} + \frac{4}{7} A^{-2} (dA)^2 - \frac{2}{7} A^{-1} \partial^i A \partial_i \Phi + \frac{2}{7} (d\Phi)^2 + \frac{c}{42 \ell} A^{-1} \slashed{Z} \Gamma_z \Gamma_{11} \right. \right.
 \\ \nonumber
 & - \frac{1}{42} A^{-1} \partial_i A \Gamma \slashed{Z}^i \Gamma_{11} + \frac{1}{21} \partial_i \Phi \Gamma \slashed{Z}^i \Gamma_{11} - \frac{1}{504} \slashed{Z} \slashed{Z} - \frac{c}{14 \ell} A^{-1} S \Gamma_z
 \\ \nonumber
 & - \frac{1}{14} A^{-1} S \slashed{\partial} A + \frac{9}{14} S \slashed{\partial} \Phi + \frac{3}{56} S \slashed{Z} \Gamma_{11} + \frac{11}{28} S^2 + \frac{c}{28 \ell} A^{-1} \slashed{F} \Gamma_z \Gamma_{11}
 \\ \nonumber
 & + \frac{1}{28} A^{-1} \partial_i A \Gamma \slashed{F}^i \Gamma_{11} + \frac{5}{28} \partial_i \Phi \Gamma \slashed{F}^i \Gamma_{11} + \frac{5}{336} \slashed{Z} \slashed{F} - \frac{1}{28} \slashed{F} \slashed{F}
 \\ \nonumber
 &  + \frac{c}{112 \ell} A^{-1} \slashed{Y} \Gamma_z + \frac{1}{112} A^{-1} \partial_i A \Gamma \slashed{Y}^i + \frac{1}{336} \partial_i \Phi \Gamma \slashed{Y}^i + \frac{1}{42 \cdot 96} \slashed{Z} \slashed{Y} \Gamma_{11}
 \\ \nonumber
 & + \frac{1}{168} S \slashed{Y} - \frac{1}{448} \slashed{F} \slashed{Y} \Gamma_{11} + \frac{1}{168 \cdot 96} \slashed{Y} \slashed{Y} + \frac{5 c}{14 \ell} A^{-1} \slashed{X} - \frac{5}{14} A^{-1} X_i \partial^i A \Gamma_z
 \\
 & \left. \left. + \frac{3}{14} X_i \partial^i \Phi \Gamma_z + \frac{3}{56} X_i \slashed{Z}^i \Gamma_z \Gamma_{11} - \frac{1}{28} X_i \Gamma \slashed{F}^i \Gamma_z \Gamma_{11} + \frac{1}{168} X_i \slashed{Y}^i \Gamma_z + \frac{1}{14} X^2 \right] \chi_+ \right\rangle~.
\end{align}

Comparing this with
\begin{align} \nonumber
 &\left\| \mathbb{B}^{(+)} \chi_+ \right\| ^2 = \left\langle \chi_+, \left[ \frac{1}{4 \ell^2} + \frac{1}{4} (dA)^2 + \frac{c}{8 \ell} A S \Gamma_z + \frac{1}{8} A S \slashed{\partial} A + \frac{1}{64} A^2 S^2 + \frac{c}{16 \ell} A \slashed{F} \Gamma_z \Gamma_{11} \right. \right.
 \\ \nonumber
 &~~+ \frac{1}{16} A \partial_i A \Gamma \slashed{F}^i \Gamma_{11} - \frac{1}{256} A^2 \slashed{F} \slashed{F} + \frac{c}{192 \ell} A \slashed{Y} \Gamma_z + \frac{1}{192} A \partial_i A  \Gamma \slashed{Y}^i
 \\ \nonumber
 &~~ + \frac{1}{768} A^2 S \slashed{Y} - \frac{1}{16 \cdot 96} A^2 \slashed{F} \slashed{Y} \Gamma_{11} + \frac{1}{192^2} A^2 \slashed{Y} \slashed{Y} + \frac{c}{8 \ell} A \slashed{X}- \frac{1}{8} A X^i \partial_i A \Gamma_z
 \\
 &~~ \left. \left.  + \frac{1}{64} A^2 X_i \Gamma \slashed{F}^i \Gamma_z \Gamma_{11} + \frac{1}{192} A^2 X_i \slashed{Y}^i \Gamma_z + \frac{1}{64} A^2 X^2 \right] \chi_+ \right\rangle~,
\end{align}
 \begin{align} \nonumber
 \left\langle \Gamma_z \mathbb{B}^{(+)} \chi_+, \mathcal{A}^{(+)} \chi_+ \right\rangle &= \left\langle \chi_+, \left[ -\frac{1}{2} \partial^i A \partial_i \Phi + \frac{c}{24 \ell} \slashed{Z} \Gamma_z \Gamma_{11} - \frac{1}{24} \partial_i A \Gamma \slashed{Z}^i \Gamma_{11} - \frac{5 c}{8 \ell} S \Gamma_z \right. \right.
 \\ \nonumber
 &~~~ - \frac{5}{8} S \slashed{\partial} A - \frac{1}{8} A S \slashed{\partial} \Phi - \frac{1}{96} A S \slashed{Z} \Gamma_{11} - \frac{5}{32} A S^2 - \frac{3c}{16 \ell} \slashed{F} \Gamma_z \Gamma_{11}
 \\ \nonumber
 & ~~~ - \frac{3}{16} \partial_i A \Gamma \slashed{F}^i \Gamma_{11} - \frac{1}{16} A \partial_i \Phi \Gamma \slashed{F}^i \Gamma_{11} - \frac{A}{192} \slashed{Z} \slashed{F} + \frac{3}{128} A \slashed{F} \slashed{F}
 \\ \nonumber
 &~~~ - \frac{c}{192 \ell} \slashed{Y} \Gamma_z - \frac{1}{192} \partial_i A \Gamma \slashed{Y}^i - \frac{1}{192} A \partial_i \Phi \Gamma \slashed{Y}^i - \frac{1}{24 \cdot 96} A \slashed{Z} \slashed{Y} \Gamma_{11}
 \\ \nonumber
 &~~~ - \frac{1}{128} A S \slashed{Y} + \frac{1}{384} A \slashed{F} \slashed{Y} \Gamma_{11} - \frac{1}{96 \cdot 192} A \slashed{Y} \slashed{Y} + \frac{c}{8 \ell} \slashed{X}
 \\ \nonumber
 &~~~ - \frac{1}{8} X^i \partial_i A \Gamma_z + \frac{1}{8} A X^i \partial_i \Phi \Gamma_z + \frac{1}{32} A X_i \slashed{Z}^i \Gamma_z \Gamma_{11}
 \\
 &~~~ \left. \left. - \frac{1}{32} A X_i \Gamma \slashed{F}^i \Gamma_z \Gamma_{11} + \frac{1}{32} A X^2 \right] \chi_+ \right\rangle~,
 \end{align}
and
 \begin{align} \nonumber
 &\left\| \mathcal{A}^{(+)} \chi_+ \right\| ^2 = \left\langle \chi_+, \left[ (d\Phi)^2 + \frac{1}{6} \partial_i \Phi \Gamma \slashed{Z}^i \Gamma_{11} - \frac{1}{144} \slashed{Z} \slashed{Z} + \frac{5}{2} S \slashed{\partial} \Phi + \frac{5}{24} S \slashed{Z} \Gamma_{11} + \frac{25}{16} S^2 \right. \right.
 \\ \nonumber
 &~~~ + \frac{3}{4} \partial_i \Phi \Gamma \slashed{F}^i \Gamma_{11} + \frac{1}{16} \slashed{Z} \slashed{F} - \frac{9}{64} \slashed{F} \slashed{F} + \frac{1}{48} \partial_i \Phi \Gamma \slashed{Y}^i
 \\ \nonumber
 &~~~ + \frac{1}{576} \slashed{Z} \slashed{Y} \Gamma_{11} + \frac{5}{192} S \slashed{Y} - \frac{1}{128} \slashed{F} \slashed{Y} \Gamma_{11} + \frac{1}{96^2} \slashed{Y} \slashed{Y} + \frac{1}{2} X^i \partial_i \Phi \Gamma_z
 \\
 &~~~ \left. \left. + \frac{1}{8} X_i \slashed{Z}^i \Gamma_z \Gamma_{11} - \frac{3}{16} X_i \Gamma \slashed{F}^i \Gamma_z \Gamma_{11} - \frac{1}{48} X_i \slashed{Y}^i \Gamma_z + \frac{1}{16} X^2 \right] \chi_+ \right\rangle~,
\end{align}
we find that
\begin{align}
 &  \nabla^2 \left\| \chi_+ \right\| ^2 + \left( 3 A^{-1} \partial_i A - 2 \partial_i \Phi \right) \nabla^i \left\| \chi_+ \right\| ^2\nonumber
 \\ \nonumber
 &= \left\| \hat{\nabla}^{(+)} \chi_+ \right\| ^2 + \frac{16}{7} A^{-2} \left\| \mathbb{B}^{(+)} \chi_+ \right\| ^2 + \frac{4}{7} A^{-1} \left\langle \Gamma_z \mathbb{B}^{(+)} \chi_+, \mathcal{A}^{(+)} \chi_+ \right\rangle + \frac{2}{7} \left\| \mathcal{A}^{(+)} \chi_+ \right\| ^2~,
\end{align}
where again $\hat{\nabla}^{( +, {1\over7}, -{1\over7} )}=\hat{\nabla}^{(+)}$.
This establishes  (\ref{maxads3}) for the massive IIA $\mathrm{AdS}_3$ backgrounds.

\section{$AdS_4$}
\label{ads4co}

To prove (\ref{maxads4}), we  expand  the third term of  (\ref{eq:massless_laplacian_expansion}) in the fields to find
\begin{align}
 & \left\langle \chi_+, \left[ -4 \Psi^{(+) i \dagger} - 2 \Gamma^i \Gamma^j \Psi^{(+)}_{j} - 12 q_1 A^{-1} \Gamma^{z i} \mathbb{B}^{(+)} - 12 q_2 \Gamma^i \mathcal{A}^{(+)} \right] \nabla_{i} \chi_+ \right\rangle
 \\ \nonumber
 &= \left\langle \chi_+, \left[ \frac{6 q_1 c}{\ell} A^{-1} \Gamma^{z i} - \left[ 3 + 6 q_1 \right] A^{-1} \partial^i A - \left[ 1 + 6 q_1 \right] A^{-1} \left( \Gamma \slashed{\partial} A \right)^i - 12 q_2 \Gamma^i \slashed{\partial} \Phi \right. \right.
 \\ \nonumber
 & \bigindent  - \frac{8 + 6 q_1 + 60 q_2}{4} S \Gamma^i - \frac{1 + 12 q_2}{4} \slashed{H}^i \Gamma_{11} - \frac{1 + 4 q_2}{4} \Gamma \slashed{H}^i \Gamma_{11}
 \\ \nonumber
 & \bigindent\left. \left. - \frac{2 + 3 q_1 + 18 q_2}{4} \Gamma^i \slashed{F} \Gamma_{11} - \frac{q_1 + 2 q_2}{16} \Gamma^i \slashed{Y} + \frac{2 - 3 q_1 + 6 q_2}{2} X \Gamma^{z x i} \right] \nabla_i \chi_+ \right\rangle~.
\end{align}
This can be written in the form $ \alpha^i \nabla_i \left\| \chi_+ \right\| ^2 + \left\langle \chi_+, \mathcal{F} \Gamma^i \tilde{\nabla}_i \chi_+ \right\rangle $ provided that   $ q_2 = -\frac{1}{6} $ and $ q_1 = \frac{1}{3} $. In particular,
\begin{align}
 & \reverseindent \left\langle \chi_+, \left[ -4 \Psi^{(+) i \dagger} - 2 \Gamma^i \Gamma^j \Psi^{(+)}_{j} - 4 A^{-1} \Gamma^{z i} \mathbb{B}^{(+)} + 2 \Gamma^i \mathcal{A}^{(+)} \right] \nabla_{i} \chi_+ \right\rangle
 \\ \nonumber
 &= \left\langle \chi_+, \left[ \frac{2 c}{\ell} A^{-1} \Gamma_{z i} - 5 A^{-1} \partial_i A - 3 A^{-1} \Gamma \slashed{\partial}_i A + 2 \Gamma_i \slashed{\partial} \Phi \right. \right.
 \\ \nonumber
 & \bigindent \bigindent \left. \left. - \frac{1}{12} \left( \slashed{\Gamma H} \right) _i \Gamma_{11} + \frac{1}{4} \slashed{H}_i \Gamma_{11} \right] \nabla^i \chi_+ \right\rangle~,
\end{align}
and so
\begin{equation}
 \mathcal{F} = \frac{2 c}{\ell} A^{-1} \Gamma_z + 3 A^{-1} \slashed{\partial} A - 2 \slashed{\partial} \Phi - \frac{1}{12} \slashed{H} \Gamma_{11}~,
\end{equation}
and $ \alpha_i = -4 A^{-1} \partial_i A + 2 \partial_i \Phi $.

Combining the ${\cal{F}}$ term with the fourth term in (\ref{eq:massless_laplacian_expansion}), we find that
\begin{align} \nonumber
 & \reverseindent \left\langle \chi_+, -2 \left( \Psi^{( + ) i \dagger} + \frac{1}{3} A^{-1} \mathbb{B}^{(+) \dagger} \Gamma^{z i} + \frac{1}{6} \mathcal{A}^{(+) \dagger} \Gamma^i + \frac{1}{2} \mathcal{F} \Gamma^i \right) \left( \Psi^{( + )}_i + \frac{1}{3} A^{-1} \Gamma_{z i} \mathbb{B}^{(+)} - \frac{1}{6} \Gamma_i \mathcal{A}^{(+)} \right) \chi_+ \right\rangle
 \\ \nonumber
 &= \left\langle \chi_+, -2 \left[ \frac{5 c}{6 \ell} A^{-1} \Gamma^{z i} + \frac{11}{6} A^{-1} \partial^i A - \frac{4}{3} A^{-1} \Gamma \slashed{\partial}^i A - \frac{5}{6} \slashed{\partial} \Phi \Gamma^i + \frac{7}{24} S \Gamma^i - \frac{1}{24} \slashed{H}^i \Gamma_{11} \right. \right.
 \\ \nonumber
 & \bigindent \bigindent \left. - \frac{1}{36} \Gamma \slashed{H}^i \Gamma_{11} + \frac{5}{48} \Gamma \slashed{F}^i \Gamma_{11} + \frac{1}{24} \slashed{F}^i \Gamma_{11} + \frac{1}{192} \Gamma \slashed{Y}^i + \frac{1}{48} \slashed{Y}^i + \frac{5}{24} X \Gamma^{z x i} \right]
 \\ \nonumber
 & \bigindent \times \left[ -\frac{c}{6 \ell} A^{-1} \Gamma_{z i} + \frac{2}{3} A^{-1} \partial_i A + \frac{1}{6} A^{-1} \Gamma \slashed{\partial}_i A - \frac{1}{6} \Gamma_i \slashed{\partial} \Phi - \frac{1}{24} S \Gamma_i + \frac{1}{12} \slashed{H}_i \Gamma_{11} \right.
 \\ \nonumber
 & \bigindent \bigindent \left. \left. - \frac{1}{72} \Gamma \slashed{H}_i \Gamma_{11} + \frac{1}{48} \Gamma \slashed{F}_i \Gamma_{11} - \frac{5}{24} \slashed{F}_i \Gamma_{11} + \frac{1}{192} \Gamma \slashed{Y}_i - \frac{1}{48} \slashed{Y}_i - \frac{1}{24} X \Gamma_{z x i} \right] \chi_+ \right\rangle
 \\ \nonumber
 &= \left\langle \chi_+, \left[ -\frac{5}{3 \ell^2} A^{-2} - \frac{14}{3} A^{-2} (dA)^2 + \frac{16}{3} A^{-1} \partial_i A \partial^i \Phi - \frac{5}{3} (d\Phi)^2 - \frac{c}{6 \ell} A^{-1} S \Gamma_z \right. \right.
 \\ \nonumber
 & \bigindent - \frac{1}{6} A^{-1} \slashed{\partial} A S + \frac{1}{6} S \slashed{\partial} \Phi + \frac{7}{48} S^2 + \frac{c}{18 \ell} A^{-1} \slashed{H} \Gamma_z \Gamma_{11} - \frac{1}{18} A^{-1} \partial_i A \Gamma \slashed{H}^i \Gamma_{11}
 \\ \nonumber
 & \bigindent + \frac{1}{18} \partial_i \Phi \Gamma \slashed{H}^i \Gamma_{11} - \frac{1}{36} S \slashed{H} \Gamma_{11} - \frac{1}{432} \slashed{H} \slashed{H} - \frac{1}{24} H^2 + \frac{c}{12 \ell} A^{-1} \slashed{F} \Gamma_z \Gamma_{11}
 \\ \nonumber
 & \bigindent + \frac{1}{12} A^{-1} \partial_i A \Gamma \slashed{F}^i \Gamma_{11} - \frac{1}{12} \partial_i \Phi \Gamma \slashed{F}^i - \frac{1}{36} \slashed{H} \slashed{F} + \frac{1}{8} \slashed{H}^i \slashed{F}_i - \frac{7}{192} \slashed{F} \slashed{F} - \frac{1}{8} F^2
 \\ \nonumber
 & \bigindent + \frac{c}{48 \ell} A^{-1} \slashed{Y} \Gamma_z + \frac{1}{48} A^{-1} \partial_i A \Gamma \slashed{Y}^i - \frac{1}{48} \partial_i \Phi \Gamma \slashed{Y}^i + \frac{1}{192} S \slashed{Y} - \frac{1}{384} \slashed{F} \slashed{Y} \Gamma_{11}
 \\ \label{eq:algebraic_product1}
 & \bigindent \left. \left. + \frac{1}{96^2} \slashed{Y} \slashed{Y} - \frac{1}{96} Y^2 + \frac{5 c}{6 \ell} A^{-1} X \Gamma_x + \frac{1}{48} X \slashed{F} \Gamma_{z x} \Gamma_{11} - \frac{5}{48} X^2 \right] \chi_+ \right\rangle~.
\end{align}

Using the field equations and Bianchi identities, the last term of (\ref{eq:massless_laplacian_expansion}) can be rewritten as
\begin{align} \nonumber
 & \reverseindent \left\langle \chi_+, \nabla_i \left[ -2 \Gamma^i \Gamma^j \Psi^{( + )}_j - 4 A^{-1} \Gamma^{z i} \mathbb{B}^{(+)} + 2 \Gamma^i \mathcal{A}^{(+)} \right] \chi_+ \right\rangle
 \\ \nonumber
 &= \left\langle \chi_+, \left[ -3 \nabla^2 \ln A + 2 \nabla^2 \Phi - \frac{1}{48} \slashed{dH} \Gamma_{11} + \frac{1}{2} \slashed{\partial} S + \frac{1}{12} \slashed{dF} \Gamma_{11} + \frac{1}{240} \slashed{dY} \right] \chi_+ \right\rangle
 \\ \nonumber
 &= \left\langle \chi_+, \left[ \frac{9}{\ell^2} A^{-2} + 12 A^{-2} (dA)^2 - 14 A^{-1} \partial_i A \partial^i \Phi + 4 (d\Phi)^2 - \frac{1}{6} H^2 + \frac{1}{2} S \slashed{\partial} \Phi \right. \right.
 \\ \nonumber
 & \bigindent \bigindent + \frac{1}{12} S \slashed{H} \Gamma_{11} + \frac{7}{4} S^2 + \frac{1}{4} \partial_i \Phi \Gamma \slashed{F}^i \Gamma_{11} + \frac{1}{24} \slashed{H} \slashed{F} - \frac{1}{8} \slashed{H}^i \slashed{F}_i
 \\ \label{eq:field_eqns_and_bianchis1}
 & \bigindent \bigindent \left. \left. + \frac{3}{8} F^2 + \frac{1}{48} \partial_i \Phi \Gamma \slashed{Y}^i - \frac{1}{96} Y^2 - \frac{5}{4} X^2 \right] \chi_+ \right\rangle~,
\end{align}
while the term in (\ref{eq:massless_laplacian_expansion}) involving the Ricci scalar  is
\begin{align} \nonumber
 & \reverseindent \frac{1}{2} R^{(6)} \left\| \chi_+ \right\| ^2 = \left\langle \chi_+, \left[ -\frac{6}{\ell^2} A^{-2} - 6 A^{-2} (dA)^2 + \frac{1}{48} Y^2 + \frac{3}{2} X^2 - \frac{3}{2} S^2 + \frac{5}{24} H^2 \right. \right.
 \\ \label{eq:curvature1}
 & \bigindent \bigindent \left. \left. - \frac{1}{4} F^2 + 8 A^{-1} \partial_i A \partial^i \Phi - 2 (d\Phi)^2 \right] \chi_+ \right\rangle~.
\end{align}
Using  \eqref{eq:algebraic_product1}, \eqref{eq:field_eqns_and_bianchis1}, and \eqref{eq:curvature1}, the right-hand-side of (\ref{eq:massless_laplacian_expansion}), apart from the first term
and the $\alpha^i \partial_i \parallel \chi_+ \parallel^2$ term,
can be expressed as
\begin{align} \nonumber
 & \reverseindent \left\langle \chi_+, \left[ \frac{4}{3 \ell^2} A^{-2} + \frac{4}{3} A^{-2} (dA)^2 - \frac{2}{3} A^{-1} \partial_i A \partial^i \Phi + \frac{1}{3} (d\Phi)^2 - \frac{c}{6 \ell} A^{-1} S \Gamma_z \right. \right.
 \\ \nonumber
 & - \frac{1}{6} A^{-1} \slashed{\partial} A S + \frac{2}{3} S \slashed{\partial} \Phi + \frac{19}{48} S^2 + \frac{c}{18 \ell} A^{-1} \slashed{H} \Gamma_z \Gamma_{11} - \frac{1}{18} A^{-1} \partial_i A \Gamma \slashed{H}^i \Gamma_{11}
 \\ \nonumber
 & + \frac{1}{18} \partial_i \Phi \Gamma \slashed{H}^i \Gamma_{11} + \frac{1}{18} S \slashed{H} \Gamma_{11} - \frac{1}{432} \slashed{H} \slashed{H} + \frac{c}{12 \ell} A^{-1} \slashed{F} \Gamma_z \Gamma_{11}
 \\ \nonumber
 & + \frac{1}{12} A^{-1} \partial_i A \Gamma \slashed{F}^i \Gamma_{11} + \frac{1}{6} \partial_i \Phi \Gamma \slashed{F}^i + \frac{1}{72} \slashed{H} \slashed{F} - \frac{7}{192} \slashed{F} \slashed{F}
 \\ \nonumber
 & + \frac{c}{48 \ell} A^{-1} \slashed{Y} \Gamma_z + \frac{13}{288} A^{-1} \partial_i A \Gamma \slashed{Y}^i + \frac{1}{192} S \slashed{Y} - \frac{1}{384} \slashed{F} \slashed{Y} \Gamma_{11}
 \\ \label{eq:algebraic_product}
 & \left. \left. + \frac{1}{96^2} \slashed{Y} \slashed{Y} + \frac{5 c}{6 \ell} A^{-1} X \Gamma_x + \frac{1}{48} X \slashed{F} \Gamma_{z x} \Gamma_{11} + \frac{7}{48} X^2 \right] \chi_+ \right\rangle
\end{align}

Comparing this to
\begin{align} \nonumber
 \left\| \mathbb{B}^{(+)} \chi_+ \right\| ^2 &= \left\langle \chi_+, \left[ \frac{1}{4 \ell^2} + \frac{1}{4} (dA)^2 + \frac{c}{8 \ell} A S \Gamma_z + \frac{1}{8} A \slashed{\partial} A S + \frac{1}{64} A^2 S^2 \right. \right.
 \\ \nonumber
 & \bigindent + \frac{c}{16 \ell} A \slashed{F} \Gamma_z \Gamma_{11} + \frac{1}{16} A \partial_i A \left( \slashed{\Gamma F} \right) ^i - \frac{1}{256} A^2 \slashed{F} \slashed{F}
 \\ \nonumber
 & \bigindent + \frac{c}{192 \ell} A \slashed{Y} \Gamma_z + \frac{1}{192} A \partial_i A  \Gamma \slashed{Y}^i + \frac{1}{768} A^2 S \slashed{Y}
 \\ \nonumber
 & \bigindent - \frac{1}{16 \cdot 96} \slashed{F} \slashed{Y} \Gamma_{11} + \frac{1}{192^2} A^2 \slashed{Y} \slashed{Y} + \frac{c}{8 \ell} A X \Gamma_x
 \\
 & \bigindent \left. \left. - \frac{1}{64} A^2 X \slashed{F} \Gamma_{z x} \Gamma_{11} + \frac{1}{64} A^2 X^2 \right] \chi_+ \right\rangle~,
\end{align}

\begin{align} \nonumber
 \left\langle \Gamma_z \mathbb{B}^{(+)} \chi_+, \mathcal{A}^{(+)} \chi_+ \right\rangle &= \left\langle \chi_+, \left[ -\frac{1}{2} \partial^i A \partial_i \Phi + \frac{c}{24 \ell} \slashed{H} \Gamma_z \Gamma_{11} - \frac{5 c}{8 \ell} S \Gamma_z - \frac{5}{8} \slashed{\partial} A S - \frac{1}{8} A S \slashed{\partial} \Phi \right. \right.
 \\ \nonumber
 & \bigindent - \frac{5}{32} A S^2 - \frac{1}{24} \partial_i A \left( \slashed{\Gamma H} \right) ^i \Gamma_{11} - \frac{1}{96} A S \slashed{H} \Gamma_{11} - \frac{3c}{16 \ell} \slashed{F} \Gamma_z \Gamma_{11}
 \\ \nonumber
 & \bigindent - \frac{3}{16} \partial_i A \left( \slashed{\Gamma F} \right) ^i \Gamma_{11} - \frac{1}{16} A \partial_i \Phi \left( \slashed{\Gamma F} \right) ^i \Gamma_{11} - \frac{A}{192} \slashed{H} \slashed{F}
 \\ \nonumber
 & \bigindent + \frac{3}{128} A \slashed{F} \slashed{F} - \frac{c}{192 \ell} \slashed{Y} \Gamma_z - \frac{1}{192} \partial_i A \Gamma \slashed{Y}^i - \frac{1}{192} A \partial_i \Phi \Gamma \slashed{Y}^i
 \\ \nonumber
 & \bigindent - \frac{1}{128} A S \slashed{Y} - \frac{1}{24 \cdot 96} A \slashed{H} \slashed{Y} \Gamma_{11} + \frac{1}{384} A \slashed{F} \slashed{Y} \Gamma_{11}
 \\
 & \bigindent \left. \left. - \frac{1}{96 \cdot 192} A \slashed{Y} \slashed{Y} + \frac{c}{8 \ell} X \Gamma_x + \frac{1}{32} A X \slashed{F} \Gamma_{z x} \Gamma_{11} + \frac{1}{32} A X^2 \right] \chi_+ \right\rangle~,
 \end{align}
and
\begin{align} \nonumber
 \left\| \mathcal{A}^{(+)} \chi_+ \right\| ^2 &= \left\langle \chi_+, \left[ (d\Phi)^2 + \frac{1}{6} \partial_i \Phi \left( \slashed{\Gamma H} \right) ^i \Gamma_{11} - \frac{1}{144} \slashed{H} \slashed{H} + \frac{5}{2} S \slashed{\partial} \Phi + \frac{5}{24} S \slashed{H} \Gamma_{11} \right. \right.
 \\ \nonumber
 & \bigindent + \frac{25}{16} S^2 + \frac{3}{4} \partial_i \Phi \left( \slashed{\Gamma F} \right) ^i \Gamma_{11} + \frac{1}{16} \slashed{H} \slashed{F} - \frac{9}{64} \slashed{F} \slashed{F}
 \\ \nonumber
 & \bigindent + \frac{1}{48} \partial_i \Phi \Gamma \slashed{Y}^i + \frac{5}{192} S \slashed{Y} + \frac{1}{576} \slashed{H} \slashed{Y} \Gamma_{11} - \frac{1}{128} \slashed{F} \slashed{Y} \Gamma_{11}
 \\
 & \bigindent \left. \left. + \frac{1}{96^2} \slashed{Y} \slashed{Y} + \frac{3}{16} X \slashed{F} \Gamma_{z x} \Gamma_{11} + \frac{1}{16} X^2 \right] \chi_+ \right\rangle
\end{align}
we find that
\begin{align}
 & \reverseindent \nabla^2 \left\| \chi_+ \right\| ^2 + \left( 4 A^{-1} \partial_i A - 2 \partial_i \Phi \right) \nabla^i \left\| \chi_+ \right\| ^2
 \\ \nonumber
 &= \left\| \hat{\nabla}^{(+)} \chi_+ \right\| ^2 + \frac{16}{3} A^{-2} \left\| \mathbb{B}^{(+)} \chi_+ \right\| ^2 + \frac{4}{3} A^{-1} \left\langle \Gamma_z \mathbb{B}^{(+)} \chi_+, \mathcal{A}^{(+)} \chi_+ \right\rangle + \frac{1}{3} \left\| \mathcal{A}^{(+)} \chi_+ \right\| ^2~,
\end{align}
where $\hat{\nabla}^{(+)}=\hat\nabla^{(+, {1\over3}, -{1\over6})}$.
This concludes the computation for this case.

\section{$AdS_n$ with $n>4$}
\label{adsnco}
Here we prove  (\ref{maxprii}) for the rest of the backgrounds, $n>4$.
As in the previous cases expressing the third term in (\ref{eq:massless_laplacian_expansion}) in terms of the fields, one finds
\begin{align}
 & \reverseindent \left\langle \chi_+, \left[ -4 \Psi^{(+) i \dagger} - 2 \Gamma^i \Gamma^j \Psi^{(+)}_{j} - 2 (10-n) q_1 A^{-1} \Gamma^{z i} \mathbb{B}^{(+)} - 2 (10-n) q_2 \Gamma^i \mathcal{A}^{(+)} \right] \nabla_{i} \chi_+ \right\rangle
 \nonumber
 \\ \nonumber
 &= \left\langle \chi_+, \left[ \frac{(10-n) q_1 c}{\ell} A^{-1} \Gamma^{z i} - \left[ 3 + (10-n) q_1 \right] A^{-1} \partial^i A - \left[ 1 + (10-n) q_1 \right] A^{-1} \left( \Gamma \slashed{\partial} A \right)^i  \right. \right.
 \\ \nonumber
 & \bigindent \bigindent - 2 (10-n) q_2 \Gamma^i \slashed{\partial} \Phi - \frac{12-n + (10-n) q_1 + 10 (10-n) q_2}{4} S \Gamma^i
 \\ \nonumber
 & \bigindent \bigindent - \frac{1 + 2 (10-n) q_2}{4} \slashed{H}^i \Gamma_{11} - \frac{3 + 2 (10-n) q_2}{12} \Gamma \slashed{H}^i \Gamma_{11}
 \\ \nonumber
 & \bigindent \bigindent - \frac{(8-n) + (10-n) q_1 + 6  (10-n) q_2}{8} \Gamma^i \slashed{F} \Gamma_{11}
 \\
 & \bigindent \bigindent \left. \left. + \frac{(n-4) - (10-n) q_1 - 2 (10-n) q_2}{96} \Gamma^i \slashed{G} \right] \nabla_i \chi_+ \right\rangle .
\end{align}
Writing this in the form  $ \alpha^i \nabla_i \left\| \chi_+ \right\| ^2 + \left\langle \chi_+, \mathcal{F} \Gamma^i \tilde{\nabla}_i \chi_+ \right\rangle $ fixes $ q_2 = -\frac{1}{10-n} $ and $ q_1 = \frac{n-2}{10-n} $ and one obtains
\begin{align}
 & \reverseindent \left\langle \chi_+, \left[ -4 \Psi^{(+) i \dagger} - 2 \Gamma^i \Gamma^j \Psi^{(+)}_{j} - 2 (n-2) A^{-1} \Gamma^{z i} \mathbb{B}^{(+)} + 2 \Gamma^i \mathcal{A}^{(+)} \right] \nabla_{i} \chi_+ \right\rangle
 \nonumber
 \\ \nonumber
 &= \left\langle \chi_+, \left[ \frac{(n-2) c}{\ell} A^{-1} \Gamma_{z i} - (n+1) A^{-1} \partial_i A - (n-1) A^{-1} \left( \Gamma \slashed{\partial} A \right) _i \right. \right.
 \\
 & \bigindent \bigindent \left. \left. + 2 \Gamma_i \slashed{\partial} \Phi - \frac{1}{12} \left( \slashed{\Gamma H} \right) _i \Gamma_{11} + \frac{1}{4} \slashed{H}_i \Gamma_{11} \right] \nabla^i \chi_+ \right\rangle .
\end{align}
Thus
\begin{equation}
 \mathcal{F} = \frac{(n-2) c}{\ell} A^{-1} \Gamma_z + (n-1) A^{-1} \slashed{\partial} A - 2 \slashed{\partial} \Phi - \frac{1}{12} \slashed{H} \Gamma_{11}~,
\end{equation}
and $ \alpha_i = -n A^{-1} \partial_i A + 2 \partial_i \Phi $.

Combining the ${\cal{F}}$ term with the fourth term in (\ref{eq:massless_laplacian_expansion}), we find
\begin{align} \nonumber
 & \reverseindent \left\langle \chi_+, -2 \left( \Psi^{( + ) i \dagger} + \frac{n-2}{10-n} A^{-1} \mathbb{B}^{(+) \dagger} \Gamma^{z i} + \frac{1}{10-n} \mathcal{A}^{(+) \dagger} \Gamma^i + \frac{1}{2} \mathcal{F} \Gamma^i \right) \right.
 \\ \nonumber
 & \bigindent \left. \times \left( \Psi^{( + )}_i + \frac{n-2}{10-n} A^{-1} \Gamma_{z i} \mathbb{B}^{(+)} - \frac{1}{10-n} \Gamma_i \mathcal{A}^{(+)} \right) \chi_+ \right\rangle
 \\ \nonumber
 &= \left\langle \chi_+, -2 \left[ \frac{(9-n) (n-2) c}{2 (10-n) \ell} A^{-1} \Gamma^{z i} - \frac{n^2 - 9n - 2}{2 (10-n)} A^{-1} \partial^i A + \frac{n^2 - 10n + 8}{2 (10-n)} A^{-1} \Gamma \slashed{\partial}^i A \right. \right.
 \\ \nonumber
 & \bigindent \bigindent - \frac{9-n}{10-n} \slashed{\partial} \Phi \Gamma^i + \frac{11-n}{4 (10-n)} S \Gamma^i - \frac{1}{4 (10-n)} \slashed{H}^i \Gamma_{11} - \frac{8-n}{24 (10-n)} \Gamma \slashed{H}^i \Gamma_{11}
 \\ \nonumber
 & \bigindent \bigindent \left. + \frac{9-n}{8 (10-n)} \Gamma \slashed{F}^i \Gamma_{11} + \frac{1}{4 (10-n)} \slashed{F}^i \Gamma_{11} + \frac{7-n}{96 (10-n)} \Gamma \slashed{G}^i + \frac{1}{8 (10-n)} \slashed{G}^i \right]
 \\ \nonumber
 & \bigindent \times \left[ -\frac{(n-2) c}{2 (10-n) \ell} A^{-1} \Gamma_{z i} + \frac{4}{10-n} A^{-1} \partial_i A + \frac{n-2}{2 (10-n)} A^{-1} \Gamma \slashed{\partial}_i A - \frac{1}{10-n} \Gamma_i \slashed{\partial} \Phi \right.
 \\ \nonumber
 & \bigindent \bigindent - \frac{1}{4 (10-n)} S \Gamma_i + \frac{8-n}{8 (10-n)} \slashed{H}_i \Gamma_{11} - \frac{1}{12 (10-n)} \Gamma \slashed{H}_i \Gamma_{11}
 \\ \nonumber
 & \bigindent \bigindent \left. \left. + \frac{1}{8 (10-n)} \Gamma \slashed{F}_i \Gamma_{11} - \frac{9-n}{4 (10-n)} \slashed{F}_i \Gamma_{11} + \frac{1}{32 (10-n)} \Gamma \slashed{G}_i - \frac{7-n}{24 (10-n)} \slashed{G}_i \right] \chi_+ \right\rangle
 \\ \nonumber
 &= \left\langle \chi_+, \left[ -\frac{(9-n) (n-2)^2}{2 (10-n) \ell^2} A^{-2} + \frac{n^3 - 11n^2 + 18n - 16}{2 (10-n)} A^{-2} (dA)^2 \right. \right.
 \\ \nonumber
 & \bigindent - \frac{2 (n^2 - 10n + 8)}{10-n} A^{-1} \partial^i A \partial_i \Phi - \frac{2 (9-n)}{10-n} (d\Phi)^2 - \frac{(n-2) c}{2 (10-n) \ell} A^{-1} S \Gamma_z
 \\ \nonumber
 & \bigindent - \frac{n-2}{2 (10-n)} A^{-1} \slashed{\partial} A S + \frac{1}{10-n} S \slashed{\partial} \Phi + \frac{11-n}{8 (10-n)} S^2 + \frac{(n-2) c}{6 (10-n) \ell} A^{-1} \slashed{H} \Gamma_z \Gamma_{11}
 \\ \nonumber
 & \bigindent - \frac{n-2}{6 (10-n)} A^{-1} \partial_i A \Gamma \slashed{H}^i \Gamma_{11} + \frac{1}{3 (10-n)} \partial_i \Phi \Gamma \slashed{H}^i \Gamma_{11} - \frac{8-n}{24 (10-n)} S \slashed{H} \Gamma_{11}
 \\ \nonumber
 & \bigindent - \frac{1}{72 (10-n)} \slashed{H} \slashed{H} - \frac{1}{24} H^2 + \frac{(n-2) c}{4 (10-n) \ell} A^{-1} \slashed{F} \Gamma_z \Gamma_{11} + \frac{n-2}{4 (10-n)} A^{-1} \partial_i A \Gamma \slashed{F}^i \Gamma_{11}
 \\ \nonumber
 & \bigindent - \frac{1}{2 (10-n)} \partial_i \Phi \Gamma \slashed{F}^i \Gamma_{11} - \frac{12-n}{48 (10-n)} \slashed{H} \slashed{F} + \frac{1}{8} \slashed{H}^i \slashed{F}^i - \frac{11-n}{32 (10-n)} \slashed{F} \slashed{F}
 \\ \nonumber
 & \bigindent - \frac{1}{8} F^2  + \frac{(n-2) c}{16 (10-n) \ell} A^{-1} \slashed{G} \Gamma_z + \frac{n-2}{16 (10-n)} A^{-1} \partial_i A \Gamma \slashed{G}^i
 \\ \nonumber
 & \bigindent - \frac{1}{8 (10-n)} \partial_i \Phi \Gamma \slashed{G}^i + \frac{7-n}{96 (10-n)} S \slashed{G} - \frac{n-4}{576 (10-n)} \slashed{H} \slashed{G} \Gamma_{11}
 \\ \label{eq:algebraic_product}
 & \bigindent \left. \left. + \frac{n-1}{48 \cdot 96 (10-n)} \slashed{G} \slashed{G} - \frac{1}{96} G^2 \right] \chi_+ \right\rangle~.
\end{align}

Using the field equations and Bianchi identities, the last term in (\ref{eq:massless_laplacian_expansion}) can be expressed as
\begin{align} \nonumber
 & \reverseindent \left\langle \chi_+, \nabla_i \left[ -2 \Gamma^i \Gamma^j \Psi^{( + )}_j - 2 (n-2) A^{-1} \Gamma^{z i} \mathbb{B}^{(+)} + 2 \Gamma^i \mathcal{A} \right] \chi_+ \right\rangle
 \\ \nonumber
 &= \left\langle \chi_+, \left[ -(n-1) \nabla^2 \ln A + 2 \nabla^2 \Phi + \frac{1}{2} \slashed{\partial} S - \frac{1}{48} \slashed{dH} \Gamma_{11} + \frac{1}{12} \slashed{dF} \Gamma_{11} + \frac{1}{240} \slashed{dG} \right] \chi_+ \right\rangle
 \\ \nonumber
 &= \left\langle \chi_+, \left[ \frac{(n-1)^2}{\ell^2} A^{-2} + (n-1) n A^{-2} (dA)^2 - 2 (2n-1) A^{-1} \partial^i A \partial_i \Phi + 4 (d\Phi)^2 \right. \right.
 \\ \nonumber
 & \bigindent \bigindent + \frac{1}{2} S \slashed{\partial} \Phi + \frac{11-n}{4} S^2 + \frac{1}{12} S \slashed{H} \Gamma_{11} - \frac{1}{6} H^2 + \frac{1}{4} \partial_i \Phi \Gamma \slashed{F}^i \Gamma_{11}
 \\ \label{eq:field_eqns_and_bianchis}
 & \bigindent \bigindent \left. \left. + \frac{1}{24} \slashed{H} \slashed{F} - \frac{1}{8} \slashed{H}^i \slashed{F}_i + \frac{7-n}{8} F^2 + \frac{1}{48} \partial_i \Phi \Gamma \slashed{G}^i \right] \chi_+ \right\rangle~,
\end{align}
while the term involving the Ricci scalar becomes
\begin{align} \nonumber
 & \reverseindent \frac{1}{2} R^{(10-n)} \left\| \chi_+ \right\| ^2 = \left\langle \chi_+, \left[ -\frac{(n-1) n}{2 \ell^2} A^{-2} - \frac{(n-1) n}{2} A^{-2} (dA)^2 + 2 n A^{-1} \partial^i A \partial_i \Phi - 2 (d\Phi)^2 \right. \right.
 \\ \label{eq:curvature}
 & \left. \left. - \frac{10-n}{4} S^2 + \frac{5}{24} H^2 + \frac{n-6}{8} F^2 + \frac{n-2}{96} G^2 \right] \chi_+ \right\rangle .
\end{align}
Using \eqref{eq:algebraic_product}, \eqref{eq:field_eqns_and_bianchis}, and \eqref{eq:curvature}, the right-hand-side of (\ref{eq:massless_laplacian_expansion}), apart from the first term and the $\alpha^i \partial_i \parallel \chi_+ \parallel^2$ term, is
\begin{align} \nonumber
 & \reverseindent \left\langle \chi_+, \left[ \frac{4 (n-2)}{(10-n) \ell^2} A^{-2} + \frac{4 (n-2)}{10-n} A^{-2} (dA)^2 - \frac{2 (n-2)}{10-n} A^{-1} \partial^i A \partial_i \Phi + \frac{2}{10-n} (d\Phi)^2 \right. \right.
 \\ \nonumber
 & - \frac{(n-2) c}{2 (10-n) \ell} A^{-1} S \Gamma_z - \frac{n-2}{2 (10-n)} A^{-1} \slashed{\partial} A S + \frac{9-n}{2 (10-n)} S \slashed{\partial} \Phi + \frac{31-3n}{8 (10-n)} S^2
 \\ \nonumber
 & + \frac{(n-2) c}{6 (10-n) \ell} A^{-1} \slashed{H} \Gamma_z \Gamma_{11} - \frac{n-2}{6 (10-n)} A^{-1} \partial_i A \Gamma \slashed{H}^i \Gamma_{11} + \frac{1}{3 (10-n)} \partial_i \Phi \Gamma \slashed{H}^i \Gamma_{11}
 \\ \nonumber
 & + \frac{12-n}{24 (10-n)} S \slashed{H} \Gamma_{11} - \frac{1}{72 (10-n)} \slashed{H} \slashed{H} + \frac{(n-2) c}{4 (10-n) \ell} A^{-1} \slashed{F} \Gamma_z \Gamma_{11}
 \\ \nonumber
 & + \frac{n-2}{4 (10-n)} A^{-1} \partial_i A \Gamma \slashed{F}^i \Gamma_{11} + \frac{8-n}{4 (10-n)} \partial_i \Phi \Gamma \slashed{F}^i \Gamma_{11} + \frac{8-n}{48 (10-n)} \slashed{H} \slashed{F}
 \\ \nonumber
 & - \frac{11-n}{32 (10-n)} \slashed{F} \slashed{F} + \frac{(n-2) c}{16 (10-n) \ell} A^{-1} \slashed{G} \Gamma_z + \frac{n-2}{16 (10-n)} A^{-1} \partial_i A \Gamma \slashed{G}^i
 \\
 & \left. \left. - \frac{n-4}{48 (10-n)} \partial_i \Phi \Gamma \slashed{G}^i + \frac{7-n}{96 (10-n)} S \slashed{G} - \frac{n-4}{576 (10-n)} \slashed{H} \slashed{G} \Gamma_{11} + \frac{n-1}{48 \cdot 96 (10-n)} \slashed{G} \slashed{G} \right] \chi_+ \right\rangle
\end{align}
Comparing this to
\begin{align} \nonumber
 \left\| \mathbb{B}^{(+)} \chi_+ \right\| ^2 &= \left\langle \chi_+, \left[ \frac{1}{4 \ell^2} + \frac{1}{4} (dA)^2 + \frac{c}{8 \ell} A S \Gamma_z + \frac{1}{8} A \slashed{\partial} A S + \frac{1}{64} A^2 S^2 \right. \right.
 \\ \nonumber
 & \bigindent + \frac{c}{16 \ell} A \slashed{F} \Gamma_z \Gamma_{11} + \frac{1}{16} A \partial_i A \left( \slashed{\Gamma F} \right) ^i - \frac{1}{256} A^2 \slashed{F} \slashed{F}
 \\ \nonumber
 & \bigindent + \frac{c}{192 \ell} A \slashed{G} \Gamma_z + \frac{1}{192} A \partial_i A  \Gamma \slashed{G}^i + \frac{1}{768} A^2 S \slashed{G}
 \\
 & \bigindent \left. \left. - \frac{1}{16 \cdot 96} \slashed{F} \slashed{G} \Gamma_{11} + \frac{1}{192^2} A^2 \slashed{G} \slashed{G}\right] \chi_+ \right\rangle~,
 \end{align}

 \begin{align} \nonumber
 \left\langle \Gamma_z \mathbb{B}^{(+)} \chi_+, \mathcal{A}^{(+)} \chi_+ \right\rangle &= \left\langle \chi_+, \left[ -\frac{1}{2} \partial^i A \partial_i \Phi + \frac{c}{24 \ell} \slashed{H} \Gamma_z \Gamma_{11} - \frac{5 c}{8 \ell} S \Gamma_z - \frac{5}{8} \slashed{\partial} A S - \frac{1}{8} A S \slashed{\partial} \Phi \right. \right.
 \\ \nonumber
 & \bigindent - \frac{5}{32} A S^2 - \frac{1}{24} \partial_i A \left( \slashed{\Gamma H} \right) ^i \Gamma_{11} - \frac{1}{96} A S \slashed{H} \Gamma_{11}
 \\ \nonumber
 & \bigindent - \frac{3c}{16 \ell} \slashed{F} \Gamma_z \Gamma_{11} - \frac{3}{16} \partial_i A \left( \slashed{\Gamma F} \right) ^i \Gamma_{11} - \frac{1}{16} A \partial_i \Phi \left( \slashed{\Gamma F} \right) ^i \Gamma_{11}
 \\ \nonumber
 & \bigindent - \frac{A}{192} \slashed{H} \slashed{F} + \frac{3}{128} A \slashed{F} \slashed{F} - \frac{c}{192 \ell} \slashed{G} \Gamma_z - \frac{1}{192} \partial_i A \Gamma \slashed{G}^i
 \\ \nonumber
 & \bigindent - \frac{1}{192} A \partial_i \Phi \Gamma \slashed{G}^i - \frac{1}{128} A S \slashed{G} - \frac{1}{24 \cdot 96} A \slashed{H} \slashed{G} \Gamma_{11}
 \\
 & \bigindent \left. \left. + \frac{1}{384} A \slashed{F} \slashed{G} \Gamma_{11} - \frac{1}{96 \cdot 192} A \slashed{G} \slashed{G} \right] \chi_+ \right\rangle~,
  \end{align}
and
 \begin{align} \nonumber
 \left\| \mathcal{A}^{(+)} \chi_+ \right\| ^2 &= \left\langle \chi_+, \left[ (d\Phi)^2 + \frac{1}{6} \partial_i \Phi \left( \slashed{\Gamma H} \right) ^i \Gamma_{11} - \frac{1}{144} \slashed{H} \slashed{H} + \frac{5}{2} S \slashed{\partial} \Phi + \frac{5}{24} S \slashed{H} \Gamma_{11} \right. \right.
 \\ \nonumber
 & \bigindent + \frac{25}{16} S^2 + \frac{3}{4} \partial_i \Phi \left( \slashed{\Gamma F} \right) ^i \Gamma_{11} + \frac{1}{16} \slashed{H} \slashed{F} - \frac{9}{64} \slashed{F} \slashed{F}
 \\
 & \bigindent \left. \left. + \frac{1}{48} \partial_i \Phi \Gamma \slashed{G}^i + \frac{5}{192} S \slashed{G} + \frac{1}{576} \slashed{H} \slashed{G} \Gamma_{11} - \frac{1}{128} \slashed{F} \slashed{G} \Gamma_{11} + \frac{1}{96^2} \slashed{G} \slashed{G} \right] \chi_+ \right\rangle~,
\end{align}
it is straightforward to see that
\begin{align}
 &  \nabla^2 \left\| \chi_+ \right\| ^2 + \left( n A^{-1} \partial_i A - 2 \partial_i \Phi \right) \nabla^i \left\| \chi_+ \right\| ^2
 \nonumber
 \\ \nonumber
 &~~= \left\| \hat{\nabla}^{(+)} \chi_+ \right\| ^2 + \frac{16 (n-2)}{10-n} A^{-2} \left\| \mathbb{B}^{(+)} \chi_+ \right\| ^2 + \frac{4 (n-2)}{10-n} A^{-1} \left\langle \Gamma_z \mathbb{B}^{(+)} \chi_+, \mathcal{A}^{(+)} \chi_+ \right\rangle
 \\
 & \bigindent
 + \frac{2}{10-n} \left\| \mathcal{A}^{(+)} \chi_+ \right\| ^2~,
\end{align}
where $\hat{\nabla}^{(+)}=\hat{\nabla}^{(+, q_1,q_2)}$ for $ q_2 = -\frac{1}{10-n} $ and $ q_1 = \frac{n-2}{10-n} $. This completes the proof of
(\ref{maxprii}) for all remaining cases.

\end{document}